\documentclass[12pt,preprinnt,ams, nofootinbib]{revtex4-1}
\usepackage{amsmath}
\usepackage{amsfonts}
\usepackage{amsthm}
\usepackage{amssymb}
\usepackage{graphicx}
\usepackage{epsfig}
\usepackage{hyperref}
\usepackage{amscd}
\usepackage{xcolor}
\usepackage{enumitem}
\usepackage{thmtools, thm-restate}
\usepackage{graphics,epsfig,placeins,subfig,wrapfig}

\newtheorem{theorem}{Theorem}[section]

\def\be{\begin{equation}}
\def\ee{\end{equation}}

\def\bea{\begin{eqnarray}}
\def\eea{\end{eqnarray}}

\def\baa{\begin{align}}
\def\eaa{\end{align}}

\theoremstyle{definition}

\theoremstyle{Identity}

\begin{document}
\preprint{APS/123-QED}
\title{Darmois matching and $C^3$ matching }

\author{Antonio  C. Guti\'errez-Pi\~{n}eres}
\email{acgutier@uis.edu.co}
\affiliation{Escuela de F\'\i sica, Universidad Industrial de Santander, CP 680002,  Bucaramanga, Colombia}
\affiliation{Instituto de Ciencias Nucleares, Universidad Nacional Aut\'onoma de M\'exico,
 AP 70543,  M\'exico, DF 04510, M\'exico}

\author{Hernando Quevedo}
\email{quevedo@nucleares.unam.mx}
\affiliation{Instituto de Ciencias Nucleares, Universidad Nacional Aut\'onoma de M\'exico, AP 70543, Ciudad de M\'exico 04510, Mexico}
\affiliation{Dipartimento di Fisica and ICRA, Universit\`a di Roma ``Sapienza", I-00185, Roma, Italy}
\affiliation{Institute of Experimental and Theoretical Physics, 
	Al-Farabi Kazakh National University, Almaty 050040, Kazakhstan}

 \begin{abstract} 
 We apply the Darmois and the $C^3$ matching conditions to three different spherically symmetric spacetimes. The exterior spacetime is described by the Schwarzschild vacuum solution whereas for the interior counterpart we choose different perfect fluid solutions with the same symmetry. We show that Darmois matching conditions are satisfied in all the three cases whereas the $C^3$ conditions are not fulfilled. We argue that this difference is due to a non-physical behavior of the pressure on the matching surface.

 \end{abstract}

\textcolor{red}{}

\maketitle

\section{Introduction}
\label{sec:int}

The problem of matching two spacetimes across a surface $\Sigma$ has been investigated for a long time \cite{lake2017revisiting}. In 1927, Darmois \cite{darm27,darmois1927equations} proposed that a physically meaningful matching can be obtained by demanding that the first and second fundamental forms (induced metric and extrinsic curvature, respectively) be continuous across $\Sigma$. Later on, in 1955, Lichnerowicz \cite{lich55} proposed an alternative approach that turned out to be equivalent to the  Darmois approach by choosing the underlying coordinates appropriately. If the fundamental forms are not continuous across the matching surface, Israel proposed in \cite{israel1966singular} to ``cover" $\Sigma$ with a shell, whose energy-momentum tensor takes care of the discontinuities. More recently, in 2012, one of us proposed in \cite{quev12} the $C^3$ approach, which is completely different because it is based not upon the use of fundamental forms ($C^2$ quantities),  but upon the behavior of the curvature eigenvalues and {their} derivatives. This approach has been applied in cosmology, and  relativistic astrophysics \cite{lq12,luongo2014characterizing,lq18,gq19,glq20}. 

Whereas {the} Darmois approach demands the continuity of the first and second fundamental forms across the matching surface, which should be specified {\it a priori}, the $C^3$ approach contains a criterium that allows us to determine the location of the matching surface. Indeed, the point is that the $C^3$ approach can be used to propose an invariant definition of repulsive gravity \cite{luongo2014characterizing}. In cosmology, this definition has been shown to be very useful because it allows us to construct models that describe inflation and the observed accelerated expansion of the Universe as repulsive gravity effects \cite{lq18,glq20}. In relativistic astrophysics, the situation is different; no repulsive gravity effects have been detected so far at the astrophysical level. In the description of the gravitational field of {compact astrophysical} objects, we use this fact in the $C^3$ approach to match vacuum and non-vacuum exact solutions in such a way that no repulsive gravity effects appear. Indeed, the behavior of the curvature eigenvalues can be used to detect regions of repulsive gravity within the gravitational field of a compact object. The {spatial} derivatives of the eigenvalues indicate the exact location where repulsion sets up. {Thus}, in the $C^3$ approach, we ``cover" the repulsion region with a different spacetime  in such a way that no repulsive gravity effects can appear. 

In a previous work \cite{gq19}, we formulated in detail the $C^3$ matching approach in asymptotically flat spacetimes. For an arbitrary metric that satisfies Einstein equations with cosmological constant and energy-momentum tensor, we computed the general form of a $6\times 6$ matrix, from which the curvature eigenvalues can be derived. The approach was applied in the case of vacuum, conformally flat, and perfect-fluid spacetimes. In the present work, we continue the investigation of $C^3$ approach. The main goal is to compare the Darmois and the $C^3$ approaches in concrete examples. Indeed, we consider spherically symmetric spacetimes and find the conditions under which the vacuum Schwarzschild spacetime can be matched with exact perfect fluid solutions. We will see that the results depend on the matching approach. In fact, in the three specific cases we will consider, it turns out that according to the Darmois approach the matching is possible whereas the conditions for a $C^3$ matching are not satisfied. This result indicates that the two approaches are {entirely} different. We discuss this contradictory result and argue that the difference can be explained by considering the physical properties of the perfect fluid solutions near the matching surface. We use this result as a motivation to propose a generalization of the $C^3$ matching procedure that allows to treat the case of discontinuities across the matching hypersurface. 

This paper is organized as follows. In Sections \ref{sec:dar} and \ref{sec:c3}, we review in detail the main aspects of the Darmois and $C^3$ matching approaches, respectively.  Sec. \ref{sec:sph} is devoted to the matching of the exterior Schwarzschild metric with three different perfect fluid solutions, namely, the Tolman III, Heintzmann II, and Buchdahl I spacetimes. We compare the results of applying both the Darmois and the $C^3$ matching approaches and establish that they lead to different results due to the presence of discontinuities of the perfect fluid parameters on the matching surface. To be able to handle such cases, we propose in Sec. \ref{sec:dis} a generalization of the $C^3$ matching procedure. 
Finally, in Sec. \ref{sec:con}, we sum up our results.


\section{Darmois matching approach}
\label{sec:dar}

In this section, we present  the  fundamental  concepts   related to the notion of hypersurfaces, which are essential for the description of {the} Darmois matching approach \cite{misner2017gravitation, bernui1994study, gourgoulhon20073}.   Let $(\cal{M}, \mathbf{g})$ represent a spacetime, where $\cal{M}$ is a real smooth (i.e. $C^{\infty}$) manifold of dimension 4 and $\mathbf g$ a Lorentzian metric on $\cal M$ with  signature $(-, +, +, +)$.  We assume that  $(\cal{M}, \mathbf{g})$ is time orientable and  $\cal M$ can be represented as a  continuous family of three-dimesional hypersurfaces 
 $\Sigma$. 
We shall restrict ourselves to spacetimes of astrophysical interest. Accordingly, it will be assumed that the topology of the spacetime $\cal{M}$ is $\Sigma \times \Re$, which is possible for a broad class of spacetimes.
On each  hypersurface $\Sigma$, a  3-metric $\mathbf {\gamma}$ is induced.  The hypersurface $\Sigma$ is said to be spacelike, if $\mathbf{\gamma}$ is definite positive (signature: $+, +, +$); timelike, if $\gamma$ is Lorentzian 
 (signature: $-, +, +$); or  null, if $\mathbf {\gamma}$  is degenerate  (signature: $0, +, +$). Accordingly, each hypersurface $\Sigma$  defines a normal vector  field $\mathbf{n} \in {\cal M}$,   
 whose norm is  $\varepsilon =   -1, +1, 0$, corresponding to    a spacelike, 
 timelike, or null  hypersurface,  respectively.
 In the following, we will assume that the spacetime  $(\cal{M}, \mathbf{g})$ is globally hyperbolic and that, thus,
any
 timelike hypersurface can be globally specified by means of a spatial  coordinate 
$x^1 = \text{constant}$. Hence, on $\Sigma$,  we can introduce a vector basis $\mathbf{e}_{i} = \{ \mathbf{e}_{0}, \mathbf{e}_{2}, \mathbf{e}_{3} \}$ and  the extrinsic curvature tensor  $\mathbf{K}$  by   
                   \begin{align}
                                       \mathbf {K}_{i j}  \equiv \ - \mathbf{e}_{j} \cdot \nabla_{i}  \mathbf{n}  =  -\ {\mathbf{e}}_{j}^{\alpha}\  \mathbf{n}_{\alpha ; i}\ ,
                      \end{align}
where  ``;" denotes the usual covariant derivative,  $ \mathbf {K}_{i j} $ denotes the components of  the extrinsic curvature tensor  $\mathbf{K}$ and  ${\mathbf{e} }_{j}^{\alpha}$, $\alpha =0,1,2,3$,  the $\alpha$th component of  the  vector  
$\mathbf{e}_{j}$.

Now we assume that it is possible to foliate the spacetime $(\cal{M}, \mathbf{g})$ into a family of slices corresponding to  timelike hypersurfaces $(\Sigma |_{x^1})_{x^1  \in  \Re}$. Therefore, there exists a smooth and regular scalar field ${X}$  on $\cal{M}$  such that on each hypersurface there is a level surface of this scalar field, i.e., 
                   $$ 
                   \forall x^1 \in \Re, \   \Sigma_{x^1} \equiv \{  p \in {\cal M}, X(p)  = x^1   \} .
                   $$
Thus, the regularity character of $X$ guarantees that the hypersufaces satisfy the condition
                  $$
                  \Sigma_{x^1}  \cap \Sigma_{{\tilde x}^1}  \ne \emptyset \ \  \text{for} \  \  x^1 \ne {\tilde x}^{1} .
                  $$
Consequently, the foliation of timelike hypersurfaces $\Sigma_{x^1}$ covers $\cal M$ in such a way that 
                 $$
                 {\cal M} =  \bigcup_{x^1 \in \Re} \Sigma_{x^1} .
                 $$
We now introduce coordinates adapted to the foliation   $(\Sigma |_{x^1})_{x^1  \in  \Re}$.  On each timelike hypersurfce $ \Sigma |_{x^1}$ one can  introduce coordinates $x^{i} = (x^0, x^2, x^3)$,  being $x^0$ a temporal coordinate whereas $x^2$ and $x^3$ are spacial coordinates.

 If these coordinates vary smoothly between  any two  infinitesimally near timelike hypersurfaces, $\Sigma |_{x^1}$  and  $\Sigma |_{x^1 + \delta x^1}$,  it is always  possible to construct  a well-behaved chart of coordinates  in $\cal M$ .  Consequently, a metric tensor on two neighboring  hypersurfaces can be given in the form
                  \begin{align*}
                                    \gamma(x^0, x^1, x^2, x^3)\  \text{and} \   \gamma(x^0, x^1 + \delta x^1, x^2, x^3),
                   \end{align*}
 respectively.  Accordingly, the metric tensor on $\cal M$ can be decomposed as 
                  \begin{align}
                                   {\bf g} = \gamma_{ij} ( \operatorname{d}x^{i}  + \beta^{i}  \operatorname{d}x^{1}) \otimes ( \operatorname{d}x^{j}  
                                             +  \beta^{j}
                               \operatorname{d} x^{1}) 
                                            + N^2  \operatorname{d}x^{1} \otimes  \operatorname{d}x^1
                  \end{align}
where  $\beta^{i}(x^{\alpha})$   is  determined  by  the  relation 
                \begin{align}
                                    x^{i}(\Sigma |_{x^1 + \delta x^1}) =   x^{i}(\Sigma |_{x^1})  -  \beta^{i}(x^{\alpha}) \delta x^1 .
                  \end{align}
Here, $ x^{i}(\Sigma |_{x^1 + \delta x^1}) $  are the coordinates  of a point in $\Sigma |_{x^1 + \delta x^1}$ constructed  by a perpendicular line  from the point $x^{i}(\Sigma |_{x^1 })$  on $\Sigma |_{x^1} $ to $\Sigma |_{x^1 + \delta x^1}$. We get then  the following expressions for the metric components
                 \begin{align}
                                    \left(
                                           \begin{array}{cc}
                                                         g_{11}  &  g_{1i} \\
                                                         g_{i1}  &   g_{ij} 
                                           \end{array}
                                     \right)  = 
                                     \left(
                                             \begin{array}{ccc}
                                              \beta_i \beta^i  + N^2 &  & \beta_i \\
                                              \beta_i &   &  \gamma_{ij} 
                                              \end{array}
                                   \right) ,
                  \end{align}
 and the components of the inverse metric are given by 
                \begin{align}
                                   \left(
                                         \begin{array}{cc}
                                                     g^{11}   &   g^{1i} \\
                                                      g^{j1}    & g^{ij} 
                                        \end{array}
                                  \right)  = 
                                    \left(
              \begin{array}{ccc}
                                          1/N^2  &    &  - \beta^{i}/  N^2 \\
                           - \beta^{i}/  N^2  &   & \gamma^{i j} + \beta^{i}\beta^{j}/N^2 
              \end{array}
                                  \right) \ ,
                \end{align}
 where $\beta_i=\gamma_{ij}\beta^j$. Naturally, for a timelike hypersurface  its normal vector $\mathbf n$ satisfies  the condition $n^{\alpha} n_{\alpha} =1$; then,
                 \begin{align} \label{eq:normal_vector}
                                   n_{\alpha}  & =(0,N, 0, 0) ,\\
                                    n^{\alpha} &  = -(\beta^{0},  -1,  \beta^{2}, \beta^{3})/N .
                     \end{align}

 Darmois matching approach is considered as a $C^2$-matching and can be formulated as follows.
            \begin{theorem}
Let   $\Sigma$  be  a  3-dimensional  timelike hypersurface  spliting  the  spacetime  $({\cal M}, \mathbf{g})$ in two 4-dimensional  manifolds $({\cal M}^-, \mathbf{g}^-)$  and $({\cal M}^+, \mathbf{g}^+)$. The  metric  tensor $\mathbf g$  is  of  class $C^3$ except on  the  hypersurface $\Sigma$ and  satisfies  the  Einstein equations  in  ${\cal M}^-$ and  ${\cal M}^+$, respectively.  Then, we say that the manifolds ${\cal M}^-$  and ${\cal M}^+$ can be matched across $\Sigma$, if the following necessary and sufficient conditions are satisfied:

\noindent
1. $ {\mathbf \gamma}^{-}|_{\Sigma}  \  = \  {\mathbf \gamma}^{+}|_{\Sigma} $  \   
(i.e., $\mathbf{\gamma}$ is  continuous  across $\Sigma$) and 

\noindent
2. 
 $ {\mathbf K}^{-}|_{\Sigma}  =   {\mathbf K}^{+}|_{\Sigma} $ \
   (i.e., ${\mathbf K}$  is  continuous  across $\Sigma$).
                  \end{theorem}

Accordingly, in summa, Darmois matching formalism consists of choosing an appropriate coordinate chart on which the metric tensor and the  extrinsic curvature of the surface are continuous and match the corresponding solutions.

\section{$C^3$ matching  approach} 
\label{sec:c3}

 The $C^3$ matching approach uses as a starting point the curvature eigenvalues, whose behavior does not depend on the choice or coordinates. This method was first proposed in relativistic astrophysics in \cite{quev12} and further applied to define repulsive gravity in \cite{lq12,luongo2014characterizing}, to investigate  cosmological models in \cite{lq18,glq20}, and to study asymptotically flat spacetimes in \cite{gq19}.
 
  For a given metric, there are several equivalent methods to calculate its curvature eigenvalues \cite{stephani2009exact}. Here, we  use the Cartan formalism of differential forms to emphasize the independence from the coordinates. Thus, consider a set of differential forms $\vartheta^a$, $a=0,..., 3$ such that 
 \be
ds^2 = g_{\mu\nu} dx^\mu\otimes dx^\nu= \eta_{ab}\vartheta^a\otimes\vartheta^b\ ,
\ee
with $\eta_{ab}={\rm diag}(-1,1,1,1)$, and $\vartheta^a = e^a_{\ \mu}dx^\mu$. The first and second Cartan equations
\be
d\vartheta^a = - \omega^a_{\ b }\wedge \vartheta^b\ ,
\ee
\be
\Omega^a_{\ b} = d\omega^a_{\ b} + \omega^a_{ \ c} \wedge \omega^c_{\ b} = \frac{1}{2} R^a_{\ bcd} \vartheta^c\wedge\vartheta^d
\ee
allow us to compute the components of the Riemann curvature tensor $R_{abcd}$  in the local orthonormal frame $\vartheta^a$. 
Moreover, we  define the Ricci tensor and the scalar curvature as  $R_{ab} = R^{c}_{\ acb} $ and $R= R^{a}_{\ a} $, respectively.
Furthermore, we introduce the bivector representation that consists in defining the curvature components $R_{abcd}$ as the components of a $6\times 6$ matrix ${\bf R}_{AB}$ according to the convention proposed in \cite{misner2017gravitation} 
(Chapter 14, Section 14.1, pp. 333-334),  
which establishes the following correspondence between 
tetrad $ab$ and bivector indices $A$:
\be
01\rightarrow 1\ ,\quad 02\rightarrow 2\ ,\quad 03\rightarrow 3\ ,\quad 23\rightarrow 4\ ,\quad 31\rightarrow 5\ ,\quad 12\rightarrow 6\ .
\ee
Hence, by using  its symmetries,  the Riemann curvature tensor $R_{abcd}$ can be explicitly expressed in matrix notation as
\be
{\bf R}_{AB}=\left(
\begin{array}{cccccc}
 R_{0101} &  R_{0102}  &  R_{0103} &  R_{0123}  & R_{0131} &  R_{0112}   \\
 R_{0201} &  R_{0202}  &  R_{0203} &  R_{0223}  & R_{0231} &  R_{0212}   \\
 R_{0301} &  R_{0302}  &  R_{0303} &  R_{0323}  & R_{0331} &  R_{0312}   \\
 R_{2301} &  R_{2302}  &  R_{2303} &  R_{2323}  & R_{2331} &  R_{2312}   \\
 R_{3101} &  R_{3102}  &  R_{3103} &  R_{3123}  & R_{3131} &  R_{3112}   \\
 R_{1201} &  R_{1202}  &  R_{1203} &  R_{1223}  & R_{1231} &  R_{1212}   
\end{array}
\right),
\label{matexplicit}
\ee

Due to the symmetry $R_{abcd} = R_{cdab}$, the matrix ${\bf R}_{AB}$ is symmetric with 21 independent components. 
The algebraic Bianchi identity $R_{a[bcd]}=0$, which in bivector representation reads 
\be 
{\bf R}_{14}+{\bf R}_{25}+{\bf R}_{36}=0\ 
\ee
reduces the number of independent components to 20. Furthermore, 
Einstein's equations\footnote{ Along this work, we use geometric units such that $k= 8\pi G c^{-4}, G = c = 1.$ }
\be
R_{ab} - \frac{1}{2} R \eta_{ab}  =k\,T_{ab}\ ,
\ee
can be written explicitly in terms of the curvature components ${\bf R}_{AB}$,  resulting in a set of ten algebraic equations that relate the components of ${\bf R}_{AB}$ and $T_{ab}$. Consequently, only ten components ${\bf R}_{AB}$ are algebraic independent and can be arranged in the $6\times 6$ curvature matrix in the following way
\be \label{eq: CurvatureTensor}
{\bf R}_{AB}=\left(
\begin{array}{cc}
	{\bf M}_1 & {\bf L} \\
	{\bf L} & {\bf M}_2 \\
\end{array}
\right),
\ee
where 
$$
{\bf L} =\left(
\begin{array}{ccc}
{\bf R}_{14}  & 	{\bf R}_{15}  & {\bf R}_{16} \\
{\bf R}_{15} - k T_{03} &  {\bf R}_{25} &  {\bf R}_{26}  \\
{\bf R}_{16} + k T_{02}  &   \quad {\bf R}_{26}  - k  T_{01} & \quad - {\bf R}_{14} -	{\bf R}_{25}   \\
\end{array}
\right), 
$$ 
and 
${\bf M}_1$ and  ${\bf M}_2$ are $3\times 3$ symmetric matrices
$$
{\bf M}_1=\left(
\begin{array}{ccc}
{\bf R}_{11} &  \quad	{\bf R}_{12} & {\bf R}_{13} \\
{\bf R}_{12} & \quad {\bf R}_{22} &  {\bf R}_{23} \\
{\bf R}_{13} & \quad   {\bf R}_{23} &  \quad - {\bf R}_{11}   -	{\bf R}_{22}   {+} k \left(\frac{T}{2} +T_{00}\right)  \\
\end{array}
\right),
$$

$$
{\bf M}_2 = 
{   \left( \\
	\begin{array}{ccc}
	-{\bf R}_{11} + k \left(\frac{T}{2} +T_{00}-T_{11} \right)   &  {-} { \bf R}_{12} - k T_{12}   & - {\bf R}_{13} - k T_{13} \\
	{-} { \bf R}_{12} - k T_{12}   &   -{\bf R}_{22} + k \left(\frac{T}{2} +T_{00}-T_{22} \right)   &  - {\bf R}_{23} - k T_{23} \\
	- {\bf R}_{13} - k T_{13}    &    - {\bf R}_{23} - k T_{23}   &   {\bf R}_{11} +	{\bf R}_{22}  {-} k T_{33}   \\
	\end{array}
	\right)   },
$$     
with $T=\eta^{ab}T_{ab}$. 
This is the most general form {of} a curvature tensor that satisfies Einstein's equations with an arbitrary energy-momentum tensor. 
The eigenvalues $\lambda_n\ (n=1,\cdots,6)$ of the matrix ${\cal R}_{AB}$ are known as the curvature eigenvalues. 
One might wonder how the eigenvalues $\lambda_n$  and the components of the Riemann tensor $R_{abcd}$ are related. To clarify this point, let us consider the simplest case in which the curvature matrix ${\bf R}_{AB}$  is diagonal. Then, from the explicit form of the curvature matrix (\ref{matexplicit}), it follows that 
\be 
\lambda_1 = R_{0101}\ , \ \lambda_2 = R_{0202}\ , \ etc.
\ee
i.e., the eigenvalues coincide with the diagonal components of ${\bf R}_{AB}$. This shows that the eigenvalues are just the non-zero tetrad components of the curvature tensor. In general, the eigenvalues depend only on the tetrad components  and can be expressed as rational functions, in which the order of the polynomials depend on the number of non-zero tetrad components.

In the $C^3$ matching approach, we do not need to give the matching surface $\Sigma$ {\it a priori}; instead, it is determined by the matching radius, $r_{match} $, defined as 
\be 
r_{match} \in [r_{rep}, \infty) \ ,  \quad r_{rep} ={\rm max}\{r_l\} \  ,  
\label{rmatch}                                                                      
\ee
where $r_l$  $(l = 1,2,... )$, with $0 < r_l < \infty$,    represents the set of solutions of the equation
\be
\frac{\partial\lambda_n^+}{\partial r}\Big|_{r=r_l} = 0 \ ,
\ee
with $\lambda_n^+$ being the curvature eigenvalues of the manifold $({\cal M}^+, {\bf g}^+)$, which is assumed to be asymptotically flat, i.e., there exists a spatial coordinate $r$ such that
 \be
 \lim_{r\to\infty} {\bf g}^+ = {\bf \eta}
 \ee
 where ${\bf \eta}$ represents the Minkowski metric.

\begin{theorem}
Let $({\cal M}^-, {\bf g}^-)$ and $({\cal M}^+, {\bf g}^+)$ be an arbitrary 
and an asymptotically flat spacetime, which satisfy Einstein equations, and 
let $\lambda_n^-$ and $\lambda_n^+$ be the curvature eigenvalues of $({\cal M}^-, {\bf g}^-)$ and $  ({\cal M}^+, {\bf g}^+)$, respectively. Then, we say that ${\cal M}^-$ and 
 ${\cal M}^+$ can be matched at the surface $\Sigma$, determined by the matching radius $r_{match}$ as defined in Eq.(\ref{rmatch}), if the necessary and sufficient condition 
 \be
[\lambda_n] \equiv \lambda_n^- - \lambda_n ^+ = 0 ,\quad n=1,\cdots, 6
\ee
is satisfied.

\end{theorem}


From a pragmatical point of view, the interior region of compact objects corresponds to the spacetime $({\cal M}^-, \mathbf{g}^-)$ whereas the exterior region is described 
by  $({\cal M}^+, \mathbf{g}^+)$. 
Then, $r_l$ represent the extrema of the exterior eigenvalues $\lambda_n^+$ and the repulsion radius $r_{rep}$ corresponds to the extremum $r_l$ with the maximum value. In other words, $r_{rep}$ is the value of $r$, where the first extremum of $\lambda_n^+$ is encountered when approaching the origin of coordinates $r=0$ coming from infinity.  The spacetimes $({\cal M}^+, \mathbf{g}^+)$ and $({\cal M}^-, \mathbf{g}^-)$ can be matched at the matching radius $r_{match}$, which can be chosen at any value of $r$ located between the repulsion radius $r_{rep}$ and infinity.

One of the first applications of the eigenvalues is the Petrov classification of the Weyl tensor (i.e. the Riemann tensor in vacuum), which was proposed in 1954 and is important for the investigation of exact solutions of Einstein equations \cite{stephani2009exact}. Indeed, according to the Petrov classification, the main type of a given Weyl tensor can be I, II, D, III, N or O, depending on the number of curvature  eigenvalues  and their degeneracy. 
The application of eigenvalues in the context of relativistic astrophysics was proposed recently in \cite{quev12}. Furthermore, eigenvalues have been used to investigate repulsive effects in black holes and naked singularities \cite{lq12}, to formulate new models of dark energy \cite{lq18}, and to study the gravitational collapse of matter \cite{glq20}. The invariant character of the curvature  eigenvalues allows us to apply them in many different physical situations and configurations.


\section{Matching spherically symmetric spacetimes}
\label{sec:sph}

In this section, we will apply the Darmois and $C^3$ matching approaches to three different perfect fluid spherically symmetric  spacetimes. 

Regarding the Darmois approach, we will consider the case of 
a  normal vector $\mathbf n$ as given in Eq.(\ref{eq:normal_vector}). Then, the extrinsic curvature  $\mathbf K$ can be expressed as 
(for details, see \cite{misner2017gravitation}, page 513)
\begin{align}\label{eq: extrinsic curvature}
{\mathbf K}_{ij} = - {\mathbf n}_{j;i}  = \frac{1}{2N}\left(  N_{i, j} + N_{j,i}   - {\mathbf\gamma}_{ij,1}  - 2 N_{k}\Gamma^{k}_{ij}    \right) \ . 
\end{align}
Additionally, we restrict ourselves to spacetiems with metric 
\begin{align}\label{eq: particular_induced_metric}
{\bf g} = {\mathbf\gamma}_{ii} \operatorname{d}x^{i} \otimes  \operatorname{d}x^{i}  +  N^2  \operatorname{d}x^{1} \otimes  \operatorname{d}x^1 ,
\end{align} 
so that the  components  of $ \bf{K}$    reduce  to 
\begin{align}\label{eq: particular_extrinsic_curvature}
{\mathbf K}_{ij} = - \frac{1}{2N} {\mathbf\gamma}_{ij,1}\ ,  \quad i, j =0, 2,  3 \ .
\end{align} 
Moreover, in the interior and exterior regions, we choose spherical coordinates and metrics of the form 
\begin{align}\label{eq:Line_Element}
{\bf g } =   - G    \operatorname{d}t \otimes    \operatorname{d}t    
+ \frac{1}{F}    \operatorname{d}r \otimes    \operatorname{d}r    +  H \operatorname{d}\Omega \otimes    \operatorname{d} \Omega \, 
\end{align}
where  $   \operatorname{d}\Omega \otimes    \operatorname{d} \Omega \equiv   \operatorname{d}\theta \otimes    \operatorname{d}\theta   
+ \sin^2\theta     \operatorname{d}\phi \otimes    \operatorname{d}\phi $ and the functions  $G$, $F$, and $H$   depend  on $r$ only.  Similarly, we suppose that the conventional matter governing the
internal spacetime dynamics is a perfect fluid determined by the energy-momentum tensor  
\begin{align}\label{eq:PerfectFluid}
\mathbf T^{\alpha \beta} = (\rho + p) \mathbf V^{\alpha} \mathbf V^{ \beta} + p {\mathbf g}^{\alpha \beta}
\end{align}
where $\rho$ and $p$ are the energy density and the pressure of the fluid, respectively, and $\mathbf V$ is the velocity of the fluid, which we choose as  the comoving velocity $\mathbf V \alpha = (-1, 0,0,0). $

According to Birkhoff's theorem, the exterior spacetime must be described by the Schwarzschild metric
\begin{align}\label{eq: Schwarzschild_Solution}
{\bf g}^+ =   - \left(1 -  \frac{2 m}{r} \right)    \operatorname{d}t \otimes  \operatorname{d}t    
+ \left(1 -  \frac{2 m}{r} \right) ^{-1}   \operatorname{d}r \otimes    \operatorname{d}r    
+ r^2 \operatorname{d}\Omega \otimes    \operatorname{d} \Omega .
\end{align}  
Furthermore, we choose the matching hypersurface as a sphere of constant radius. Hence, according to equations (\ref{eq: particular_induced_metric}) and (\ref{eq: particular_extrinsic_curvature}),  the metric tensor induced on $\Sigma$ 
and its extrinsic curvature are given by
\begin{align}\label{eq:InducedMetricSchwarzschild} 
{\mathbf \gamma}^+ _{tt} = -1 +  \frac{2m}{r}\ ,
\quad {\mathbf \gamma}^+_{\theta\theta} =  r^2  \ ,
\quad {\mathbf \gamma}^+_{\phi \phi} = r^2 \sin^2{\theta} \ ,
\end{align}
and
\begin{align}\label{eq:ExtrinicCurvatureSchwarzschild} 
{\mathbf K}^+_{tt} = \frac{m}{r^2}\left( 1 -  \frac{2m}{r}\right)^{1/2}\ ,
\quad {\mathbf K}^+_{\theta\theta} = - r \left( 1 -\frac{2m}{r}  \right)^{1/2}\ , 
\quad {\mathbf K}^+_{\phi \phi} =  \sin^2{\theta}\, {\mathbf K}^+_{\theta \theta} ,
\end{align}
respectively.

For the $C^3$ approach we only need to calculate the curvature eigenvalues. We choose the orthonormal tetrad $\vartheta^a$ as
\be
\vartheta^0 = \left(1-\frac{2m}{r}\right)^{1/2} \operatorname{d} t\ , \
\vartheta^1 = \left(1-\frac{2m}{r}\right)^{-1/2} \operatorname{d} r\ , \
\vartheta^2 = r  \operatorname{d} \theta \ , \
\vartheta^3=  r\sin\theta \, \operatorname{d} \phi \ .
\ee
A straightforward computation shows that the curvature matrix 
${\bf R}_{AB}$ is diagonal and the eigeinvalues are
\begin{align}
\label{eq: Schwarzschild_Eigenvalues}
\lambda^+_2 & =   \lambda^+_3 =   - \lambda^+_5=  - \lambda^+_6 =  { m}/{r^3}, \\ 
\lambda^+_1  & =  - \lambda^+_4 =  - { 2 m}/{r^3}. 
\end{align}     
We can now perform the first step of the $C^3$ approach, which consists in finding the extrema of the exterior eigenvalues. Obviously, none of the Schwarzchild eigenvalues has an extremum. This means that there is no repulsion radius $r_{rep}$, which indicates in the approach the smallest sphere at which the matching can be carried out. Consequently, there is no repulsion region in the Schwarzschild spacetime that should be covered by an interior solution, which is the conceptual background of the $C^3$ approach. 
Then, the matching radius can be located anywhere outside the central singularity, i.e., $r_{match} \in (0,\infty)$.

\subsection{The  Tolman III spacetime} 
\label{sec:tol}
	
The Tolman III spacetime is an exact solution of Einstein equations that describes a perfect fluid with constant energy density. It can be written out as  \cite{tolman1939static, delgaty1998physical} 
\begin{align}\label{eq: Solution_1}
{\bf g}^-  & =   - \left[  \frac{3}{2} f(K)     - \frac{1}{2}f(r) \right]^2   \operatorname{d}t \otimes    \operatorname{d}t    
\nonumber\\  & +  \frac{1}{f^2(r)}    \operatorname{d}r \otimes    \operatorname{d}r   
+   r^2   \operatorname{d}\Omega \otimes    \operatorname{d} \Omega \ ,   \\    
\rho  & = \frac{3m}{4 \pi K^3}  \ ,\quad 
p       =   \frac{3m}{4\pi K^3} \frac{\left[ f(r) - f(K) \right ] }
{\left[    3  f(K) - f(r) \right ]}\ , \ \ f(r) = \left(1- \frac{2mr^2}{K^3}\right)^{1/2}  \ ,               
\end{align}
where $m$ and  $K$ are constants. Furthermore, the components of the metric on $\Sigma$ are given by 
\be
\gamma^-_{tt} = - \left[  \frac{3}{2} f(K)   - \frac{1}{2} f(r) \right]^2 \ , \ 
\gamma^-_{\theta\theta} = r^2\ ,\ 
\gamma^-_{\phi\phi} =  r^2 \sin^2 \theta \ ,
\label{KTint}
\ee
and the components of the corresponding extrinsic curvature become
\begin{align}\label{eq:ExtrinicCurvatureSol1} 
{\mathbf K}^-_{tt} & =\frac{rm}{2K^{9/2}} [ 3K(K- 2m )^{1/2} -(K^3 - 2mr^2)^{1/2}  ]\ , \nonumber\\
{\mathbf K}^-_{\theta \theta}&  = - \frac{r}{K^{3/2}} ( K^3 - 2m r^2)^{1/2}\ ,
{\mathbf K}^-_{\phi \phi} =   \sin^2{\theta}    {\mathbf K}^- _{\theta \theta} \ .
\end{align}

Consider now the matching hypersurface $\Sigma$ as a sphere of radius $r = r_0 = $ const.  Then, 
by imposing the first Darmois condition on the exterior  (\ref{KTint}) and interior   (\ref{eq:InducedMetricSchwarzschild})  metrics, it follows that 
\begin{align}{\mathbf \gamma}^+ |_{r=r_0}  \ = \  {\mathbf \gamma}^- |_{r=r_0}  \ ,
\end{align}
only if $K=r_0$. Furthermore, a comparison of the components of the extrinsic curvature  (\ref{eq:ExtrinicCurvatureSol1}) and 
(\ref{eq:ExtrinicCurvatureSchwarzschild}) leads to 
\begin{align}
{\mathbf K}^+|_{r=r_0}   =   {\mathbf K}^- |_{r=r_0}    \ ,
\end{align}
implying that the second Darmois condition is satisfied identically.  We conclude that according to Darmois approach, the exterior Schwarzschild metric and the interior Tolman III solution can be matched on the hypersurface $r=r_0=$ const. and, consequently, determine a physically meaningful spacetime. 
Notice that on the matching hypersurface $r=r_0$, the pressure vanishes, but the density remains constant, $\rho(r_0)= 3m /4\pi r_0^3$.

Consider now the $C^3$ matching approach. The choice of the differential forms $\vartheta^a$ is suggested by the 
diagonal form of the metric (\ref{eq: Solution_1}). Then,
\be 
\vartheta^0 = \left[  \frac{3}{2} f(K)  - \frac{1}{2} f(r)  \right] \operatorname{d}t \ ,
\ee
\be
\vartheta^1 = \frac{1}{f(r)}   \operatorname{d}r\ , \
\vartheta^2 =r \operatorname{d} \theta\ ,\ 
\vartheta^3 = r \operatorname{d} \phi \ .
\ee
The computation of the corresponding  matrix ${\bf R}_{AB}$ yields the following curvature eigenvalues
\begin{align}
\lambda^-_1    & =  \lambda^-_2 =  \lambda^-_3 = \frac{    2m ( K^3 - 2m r^2)^{1/2}  }{   K^3[ 3K(K - 2m )^{1/2} - (K^3 - 2mr^2)^{1/2}]    } \ ,
\end{align} 
\begin{align}   
\lambda^-_4  & =  -  \lambda^-_1   + 4\pi ( \rho  + p ) = \frac{2m}{K^3} \ ,\\
\lambda^-_5   & =  -   \lambda^-_2   + 4\pi ( \rho + p) = \frac{2m}{K^3} \ ,\\
 \lambda^-_6   & =  -     \lambda^-_3  + 4\pi ( \rho + p ) = \frac{2m}{K^3} \ .
\end{align} 
The second step of the $C^3$ matching approach implies that on the matching hypersurface, $r=r_{match}$, all the eigenvalues should coincide, $\lambda^+_n= \lambda^-_n, \ \forall\, n$. A comparison of the above expressions with the Schwarzschild eigenvalues (\ref{eq: Schwarzschild_Eigenvalues}) shows that there is no $r_{match}$ for which all the conditions are satisfied. We conclude that according to the $C^3$ matching approach, the interior Tolman III solution cannot be matched with the exterior Schwarzschild metric.


\subsection{The Heintzmann II spacetime}
\label{sec:hei}

The Heintzmann II spacetime is a perfect fluid solution of Einstein equations described by  the metric
 \cite{heintzmann1969new, delgaty1998physical} 
\begin{align} 
{\bf g}^-    &  =   - A^2 (1 + a r^2)^3  \operatorname{d}t \otimes    \operatorname{d}t    
+   \left\{1 - \frac{3 a r^2}{2 (1 + ar^2)} \left[  1 + K ( 1 + 4 a r^2)^{-1/2}\right] \right\}^{-1}
\operatorname{d}r \otimes    \operatorname{d}r   \nonumber\\
&        +   r^2   \operatorname{d}\Omega \otimes    \operatorname{d} \Omega \ , \label{eq:S2_Sol_Metric}   \\     
p  &  =    - \frac{ 3a\left[ (7ar^2 +1)K + 3(ar^2-1) ( 4ar^2 + 1)^{1/2}\right]}{16 \pi (4 a r^2 +1)^{1/2} (ar^2 + 1)^2}  , \label{eq:S2_Sol_Pressure}\\
\rho   &  =     \frac{3a \left[ 3(3ar^2 + 1)K  + (4a^2r^4 + 13ar^2 + 3) (4ar^2 +1)^{1/2}\right]}{16 \pi (4 a r^2 +1)^{3/2} (ar^2 + 1)^2}  \ ,\label{eq:S2_Sol_Energy}     
\end{align}
where  $a$, $A$ and $K$ are  constants. 

This metric tensor induces on a hypersurface $\Sigma$ with $r=$const. a 3-metric with extrinsic curvature given by
\begin{align}\label{eq:InducedMetricS2_Sol}
{\mathbf \gamma}^- _{tt} = -A^2 (ar^2 + 1)^3 , \quad 
{\mathbf \gamma}^- _{\theta \theta}  =  r^2 \ , \quad 
 {\mathbf \gamma}^- _{\phi \phi}  =  r^2 \sin^2\theta \ , 
\end{align}
and
\begin{align}\label{eq:ExtrinicCurvatureS2_Sol} 
{\mathbf K}^-_{tt}                  & =   - \frac{3\sqrt{2} A^2 a r( ar^2 + 1)^{3/2} }{2(4ar^2 + 1)^{1/4}}  \left[ (2 - ar^2)(4ar^2 + 1)^{1/2} - 3ar^2K\right]^{1/2}\ , \nonumber \\
{\mathbf K}^-_{\theta \theta} & =   \frac{\sqrt{2}r}{2(ar^2 + 1)^{1/2} (4 ar^2 + 1)^{1/4}}  \left[ (2 - ar^2)(4ar^2 + 1)^{1/2} - 3ar^2K\right]^{1/2}\  \nonumber, \\
{\mathbf K}^- _{\phi \phi}       & =   \sin^2{\theta}    {\mathbf K}^- _{\theta \theta} \ .
\end{align}            

We now consider as matching hypersurface a sphere of radius $r= r_0$, where we impose the Darmois conditions ${\mathbf \gamma}^- |_{r=r_0 }    =    {\mathbf \gamma}^+ |_{r=r_0 }$ and  ${\mathbf K}^- |_{r=r_0}  =   {\mathbf K}^+ |_{r= r_0}$. Lengthy calculations show that in this case Darmois conditions are equivalent to fixing the arbitrary constants that enter the spacetime metric as
\begin{align}\label{eq:Sol2_Parameters}
a =   -\frac{m}{(7m -3 r_0)r_0^2}\ , A=  -\frac{3^{1/2} (3r_0 - 7m)^{3/2}}{r_0^{1/2} (18m - 9r_0)}\ ,
K = -\frac{3^{1/2}(8m - 3r_0)(m - r_0)^{1/2} }{r_0(7m - 3r_0)} ,
\end{align}
in terms of the Schwarzschild mass $m$ and the radius of the matching sphere $r_0$. We conclude that according to Darmois approach the Heintzmann II spacetime can be matched with the exterior Schwarzschild metric. 

\begin{figure}[h!] 
	\centering
	\subfloat[$p$]{ \includegraphics[width = 0.45\textwidth]{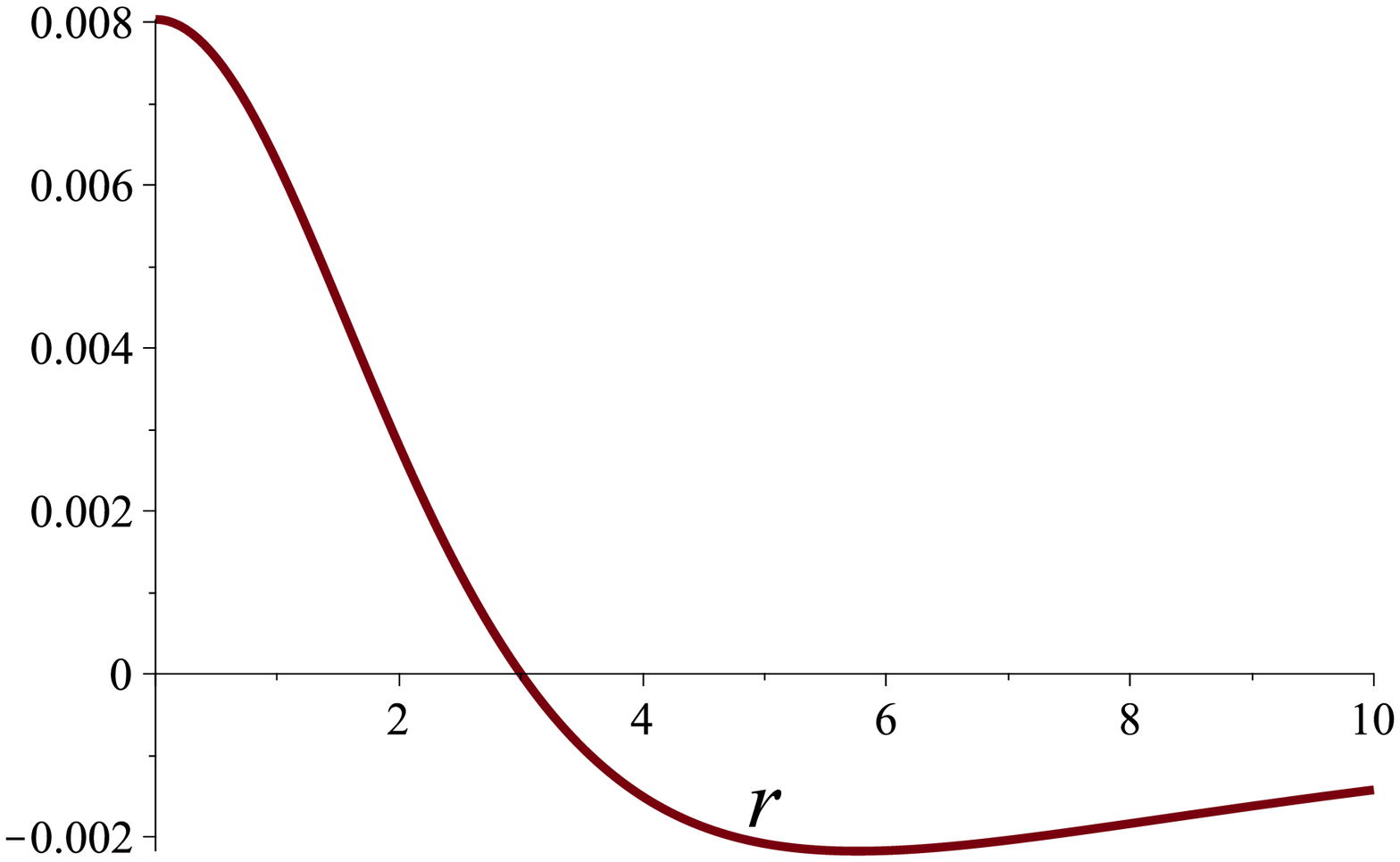}}
	\subfloat[$\rho$]{\includegraphics[width = 0.45\textwidth]{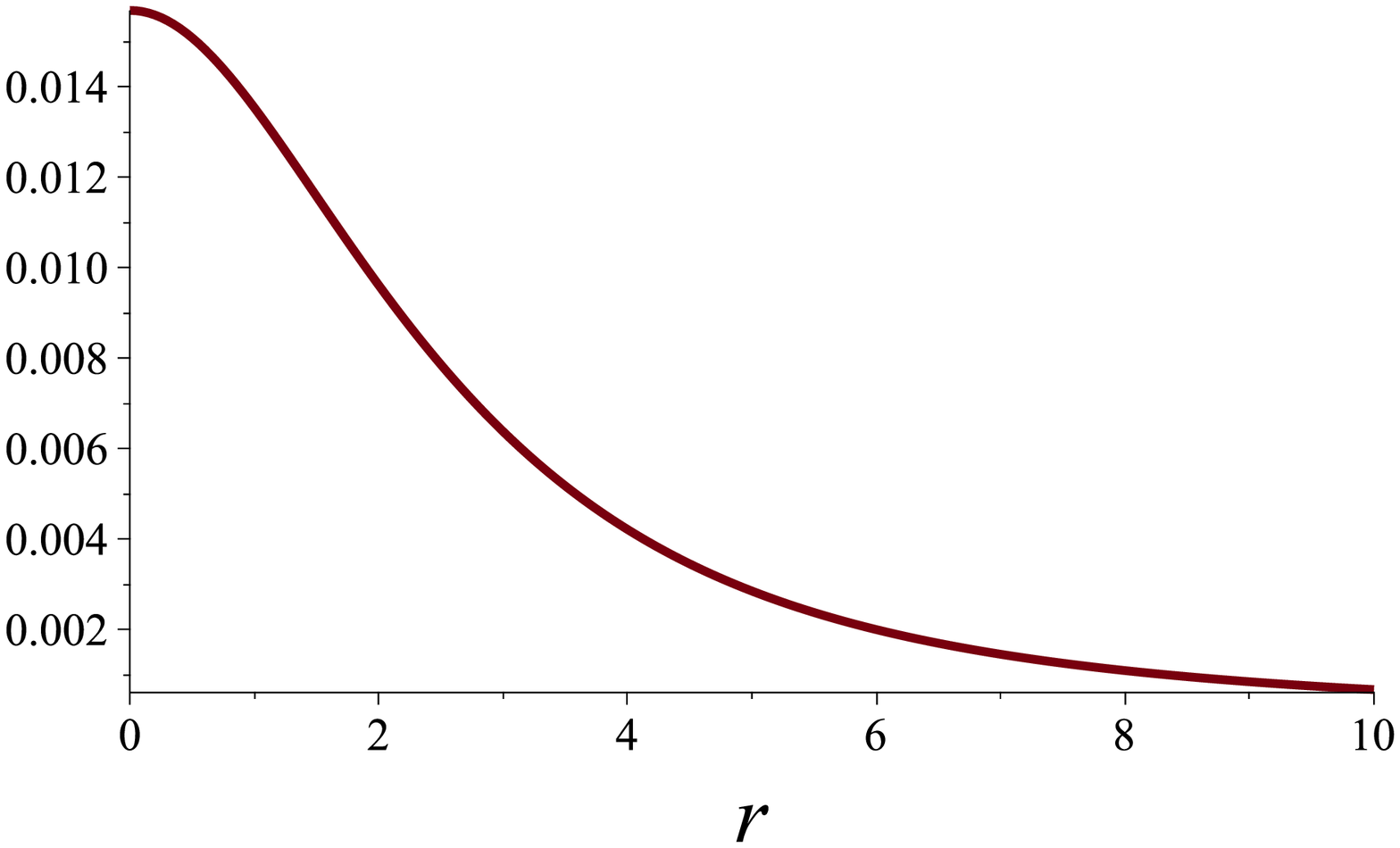}}
	\caption{ Pressure and the energy density of the Heintzmann II solution for $r_0 =3$ and $m=1$.}
	\label{fig:Sol2_Presure_and_Energy} 
\end{figure}

We notice that if we introduce the above values for the constants $a$, $A$, and $K$ into the expressions for the pressure and density of the Heintzmann solution, we obtain that on the matching hypersurface, the pressure vanishes,  but the density is different from zero. In fact, the density tends to zero only asymptotically. We illustrate this behavior in Fig.  \ref{fig:Sol2_Presure_and_Energy}, where for concreteness we set $r_0=3$. We see that the pressure is positive inside the source, $0<r_0<3$, vanishes on the matching hypersurface, $r_0=3$, and becomes negative outside the body. On the other hand, the density is always positive and vanishes only asymptotically.

\begin{figure}[h!]
	\centering
	\subfloat[$\lambda_1$]{  \includegraphics[width = 0.5\textwidth]{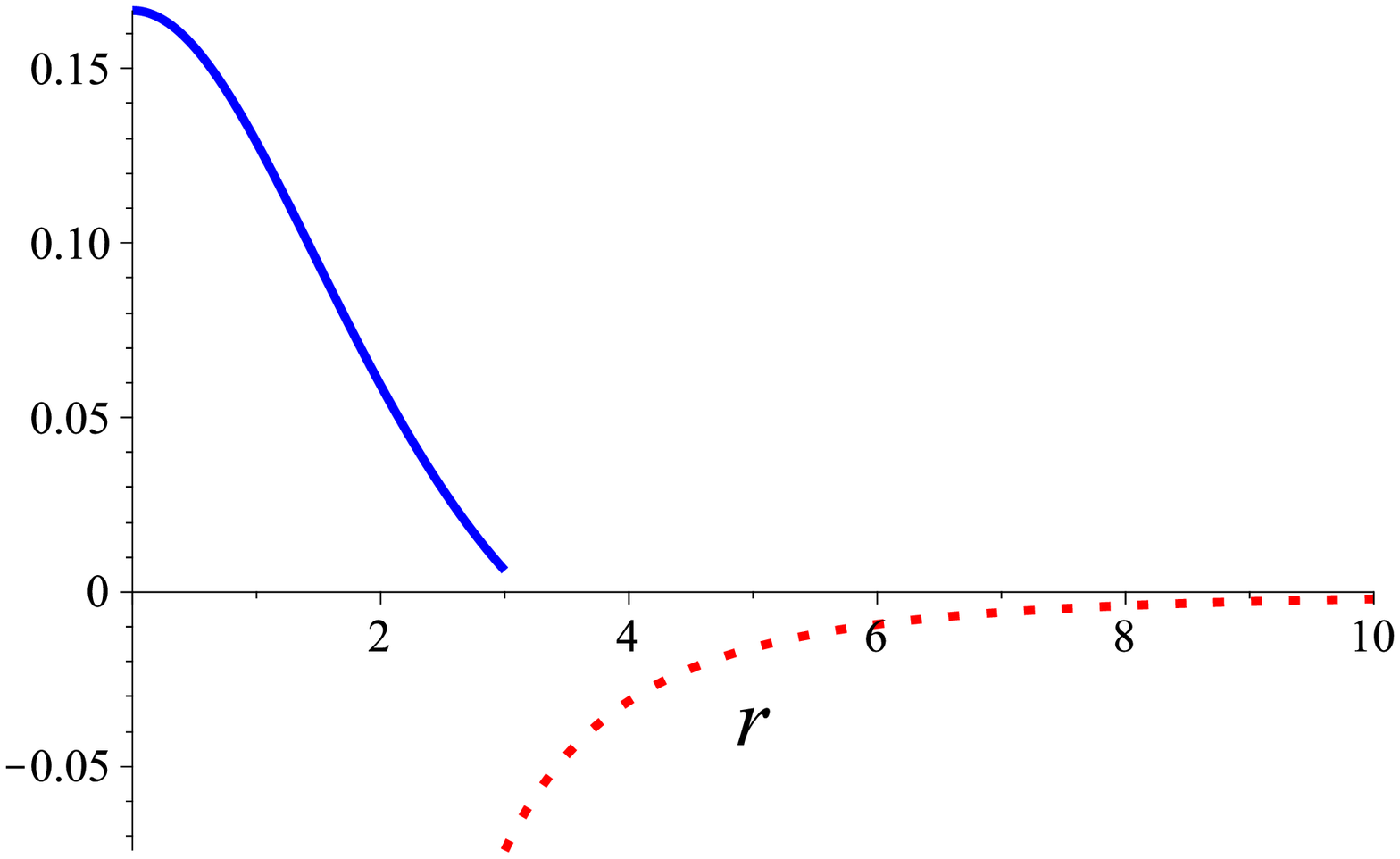} }
	\subfloat[$\lambda_2$]{  \includegraphics[width = 0.5\textwidth]{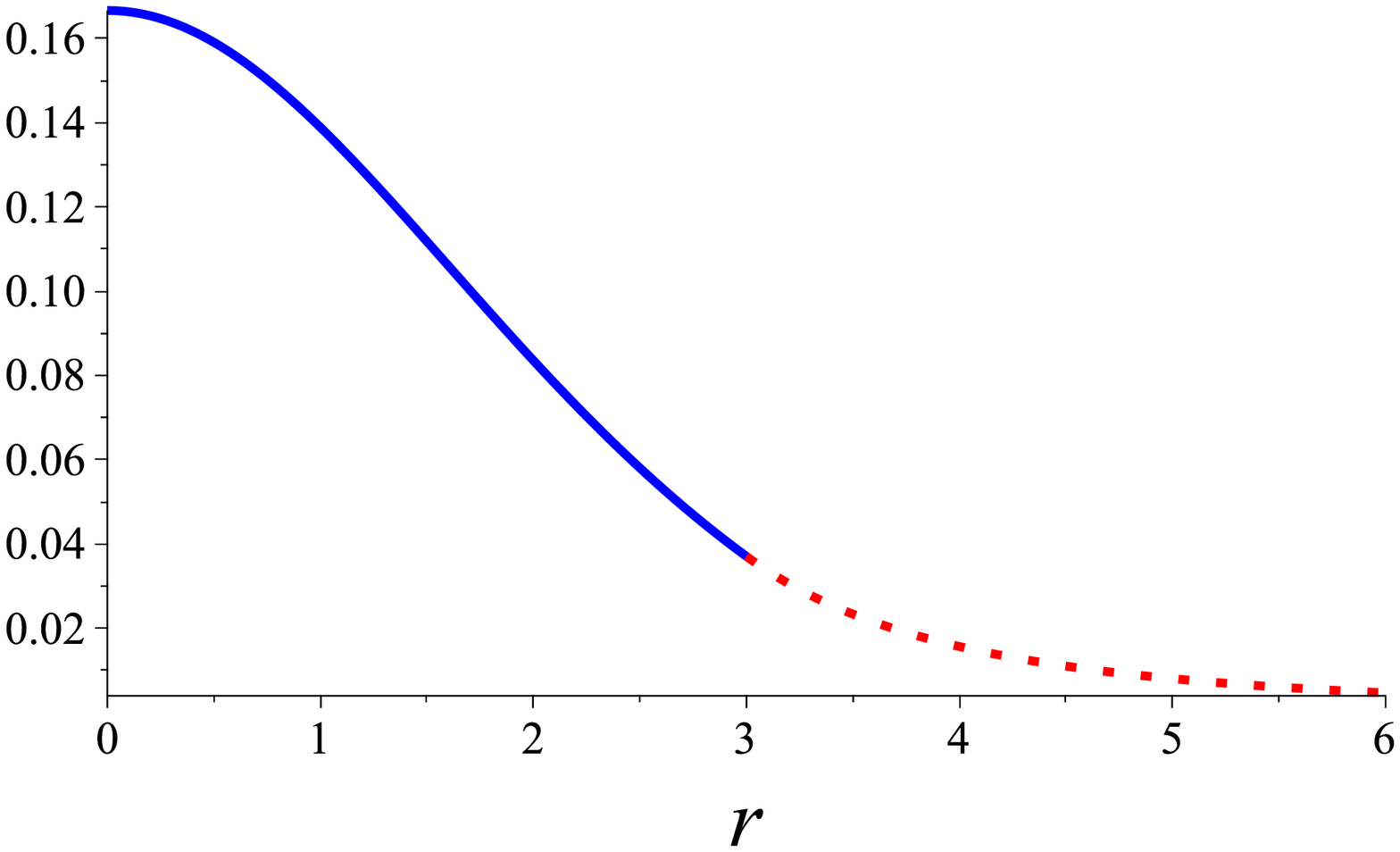} } \\
	\subfloat[$\lambda_3$]{  \includegraphics[width = 0.5\textwidth]{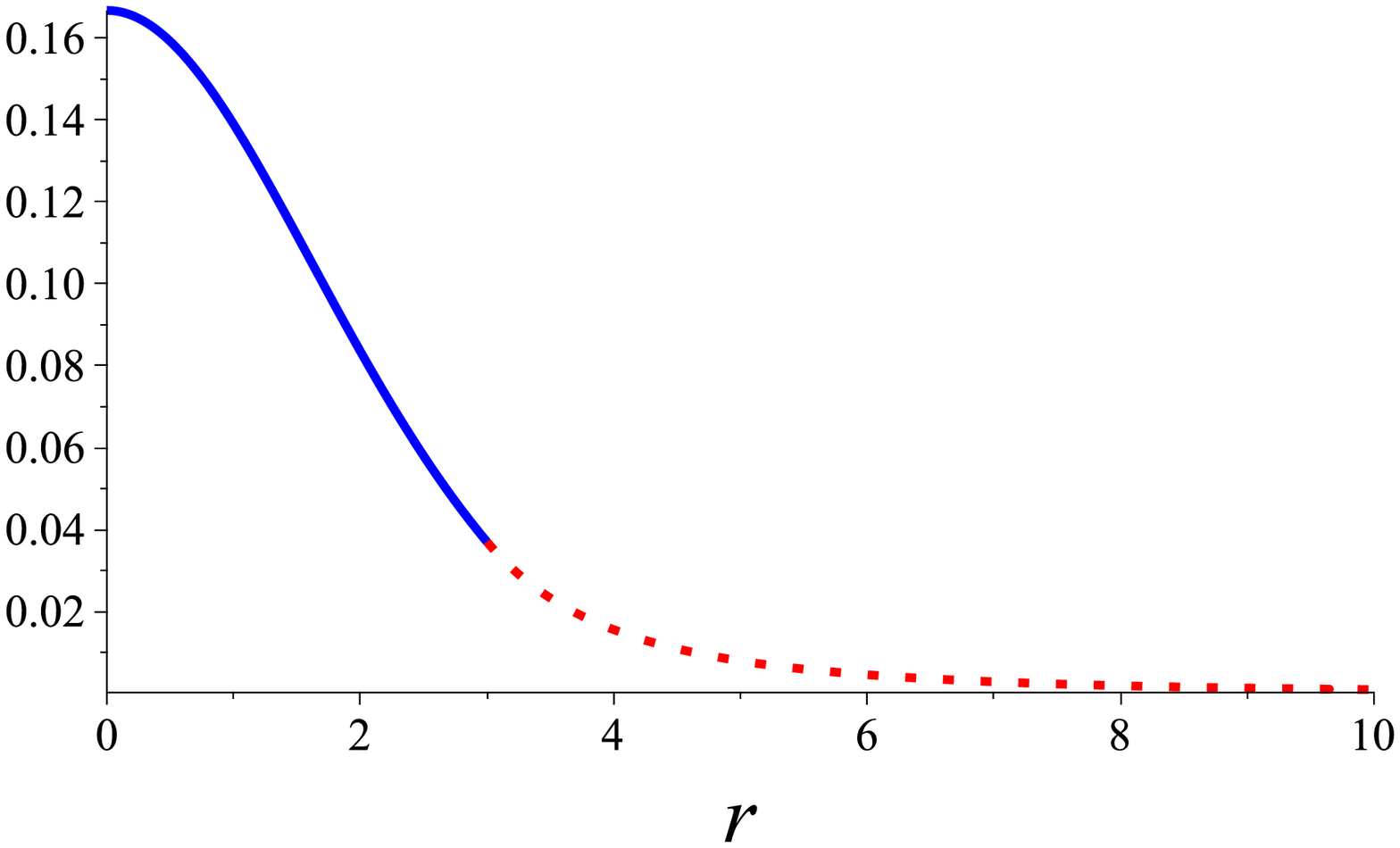} } 
	\subfloat[$\lambda_4$]{  \includegraphics[width = 0.5\textwidth]{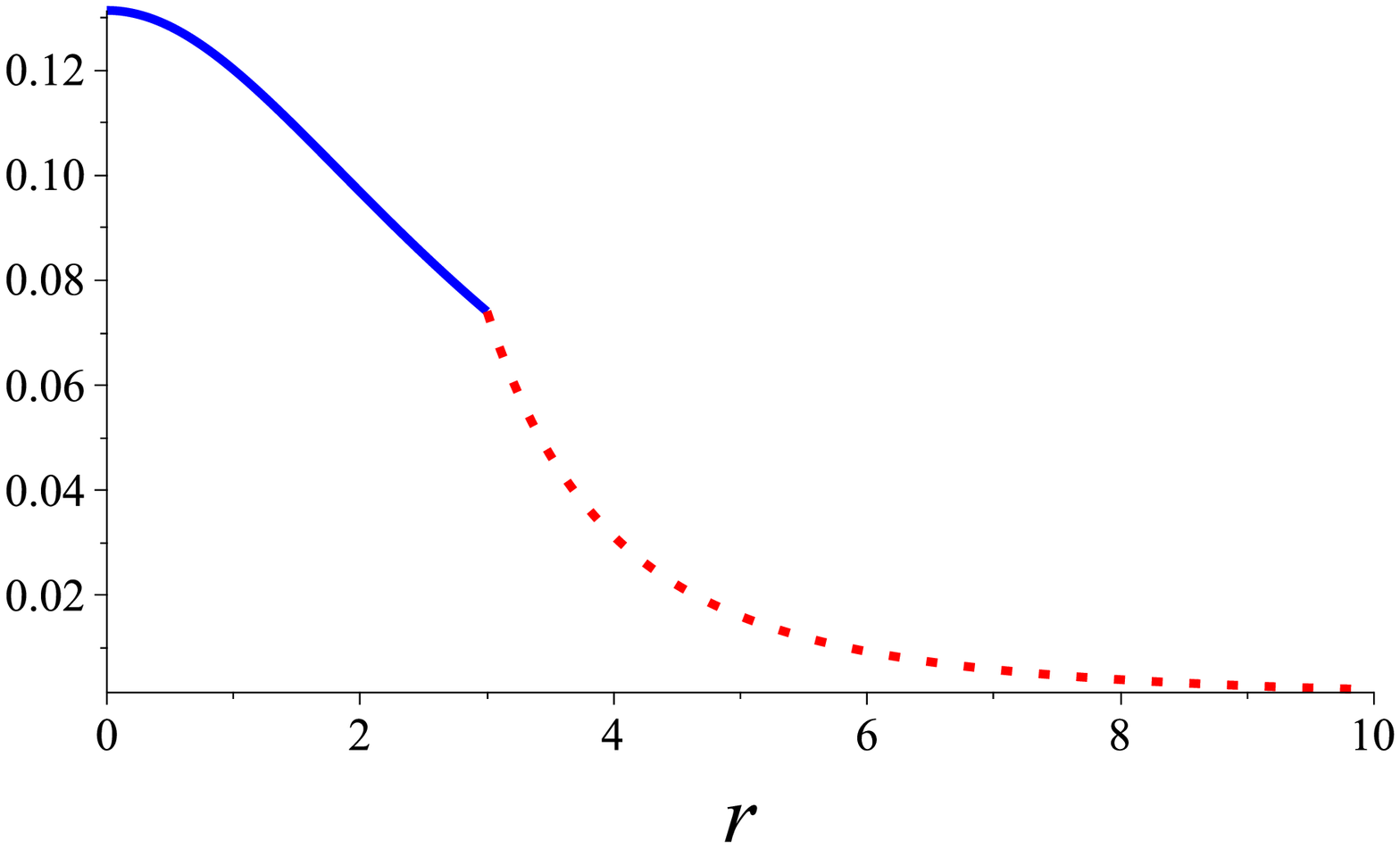} } \\
	\subfloat[$\lambda_5$]{  \includegraphics[width = 0.5\textwidth]{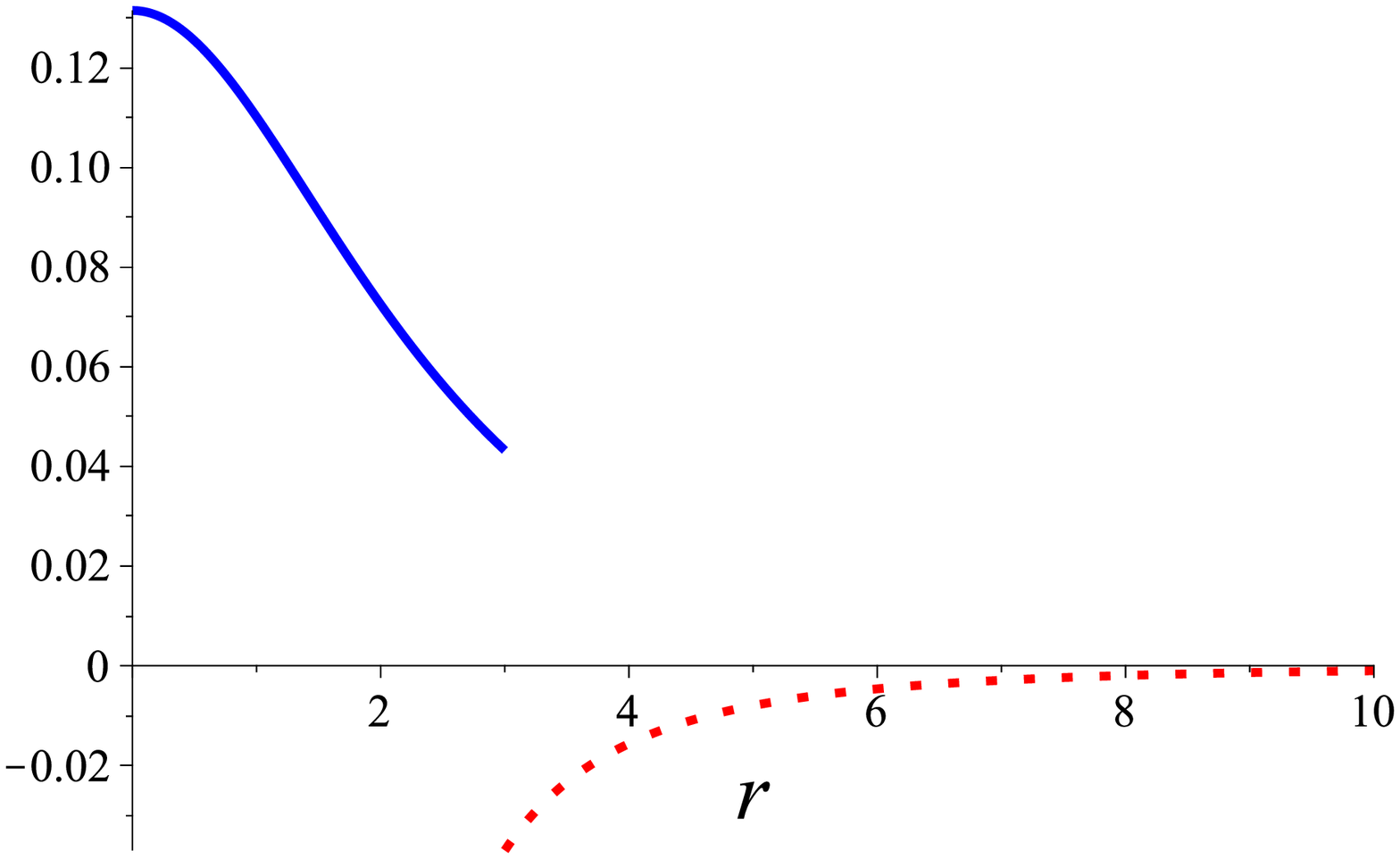} }
	\subfloat[$\lambda_6$]{  \includegraphics[width = 0.5\textwidth]{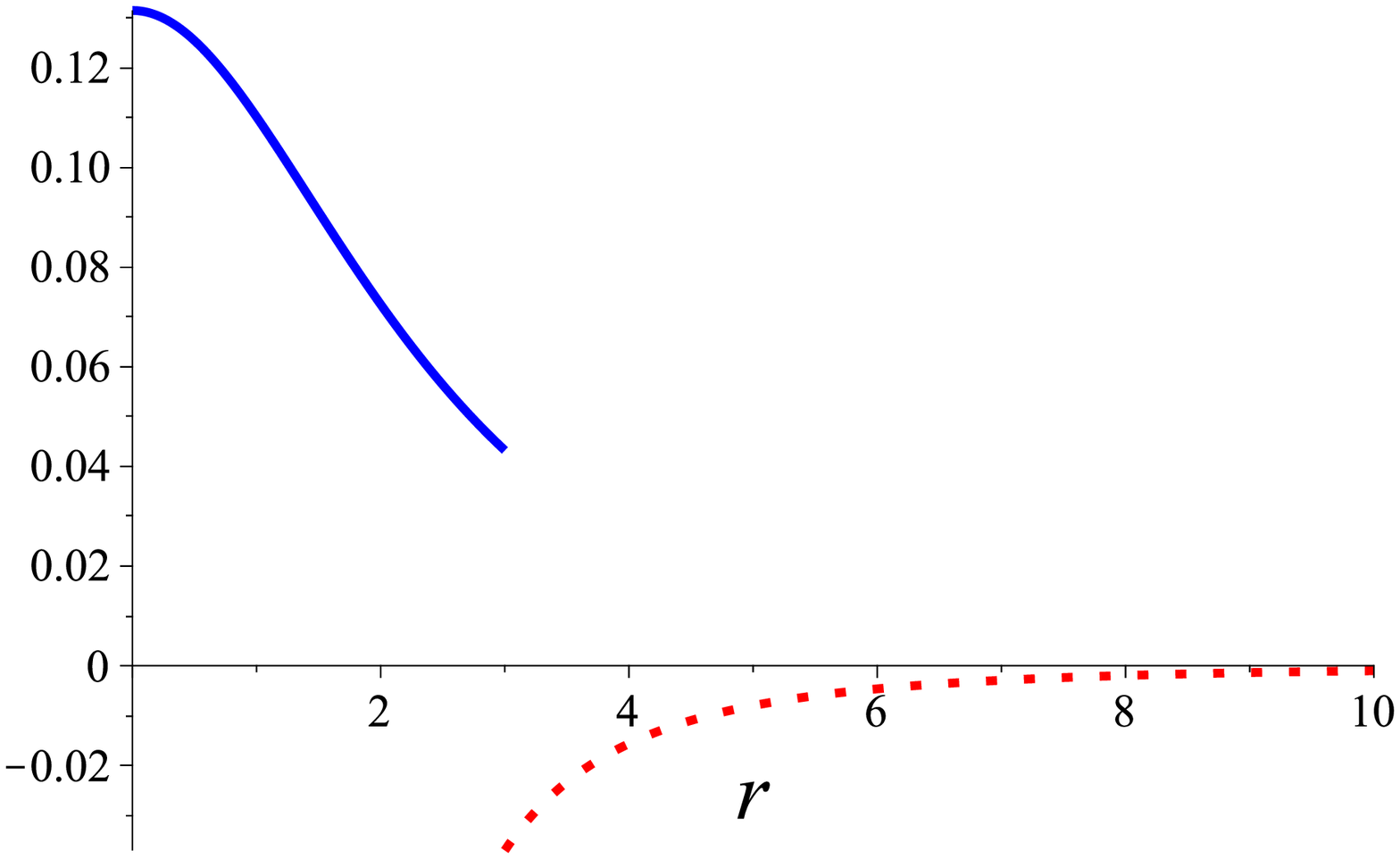} }
         \caption{ The Eigenvalues of the Heintzmann II metric ($r\leq 3$) and of the Schwarzschild metric $(r\geq 3$) for $r_0 = 3$ and  $m=1$. }
\label{fig:Sol2_eigenvalues}
\end{figure}

To apply the $C^3$ approach, we choose the orthonormal tetrad as 
\be 
\vartheta^0 = A (1 + a r^2)^{3/2}\operatorname{d}t \ ,\ 
\vartheta^2 =r \operatorname{d} \theta\ ,\ 
\vartheta^3 = r \operatorname{d} \phi \ ,
\ee
\be
\vartheta^1 = \left\{1 - \frac{3 a r^2}{2 (1 + ar^2)} \left[  1 + K ( 1 + 4 a r^2)^{-1/2}\right] \right\} ^{-1/2}   \operatorname{d}r\ ,
\ee
which leads to the curvature eigenvalues
\begin{align}\label{eq:Sol2_lambda(1)}
\lambda^-_1 = -\frac{3a [3aKr^2(3ar^2 + 1) + (4ar^2 + 1)^{1/2}( 4a^2r^4 - 3ar^2 -1) ]}{ (4ar^2 + 1)^{3/2} (ar^2 + 1)^2} \ ,
\end{align}
\begin{align}\label{eq:Sol2_lambda(2-3)}
\lambda^-_2 =   \lambda^-_3 = - \frac{ 3a[  (ar^2 - 2)(4ar^2+ 1)^{1/2} + 3aKr^2 ]}{2(4ar^2 + 1)^{1/2}(ar^2 + 1)^2} \ ,
\end{align}
\begin{align}\label{eq:Sol2_lambda(4)}
\lambda^-_4 =  -\lambda^-_1 + 4\pi ( \rho + p )  = \frac{3a[(4ar^2 + 1)^{1/2}  + K]}{ 2(4ar^2 + 1)^{1/2}( ar^2 +1 )} \ ,
\end{align}
\begin{align}\label{eq:Sol2_lambda(5)}
\lambda^-_5  & =  -\lambda^-_2 + 4\pi ( \rho+ p )   \nonumber\\
& = -\frac{3a [ 2Kar^2(ar^2 -1) -(4ar^2 +1)(4ar^2 +1)^{1/2} - K]}{ 2(4ar^2 + 1)^{3/2} (ar^2 + 1)^2} \ ,
\end{align}
\begin{align}\label{eq:Sol2_lambda(6)}
\lambda^-_6 & =  -\lambda^-_3 + 4\pi ( \rho + p )  =   \nonumber\\ \ 
& = -\frac{3a [ 2Kar^2(ar^2 -1) -(4ar^2 +1)(4ar^2 +1)^{1/2} - K]}{ 2(4ar^2 + 1)^{3/2} (ar^2 + 1)^2}  \ .
\end{align}
It is then possible to prove that there is no solution to the $C^3$ matching condition, which implies the equality between the interior and exterior eigenvalues. This incompatibility is illustrated in Fig.  \ref{fig:Sol2_eigenvalues}, where we plot the  eigenvalues of the Heintzmann II metric for $a$, $A$, and  $K$ given by (\ref{eq:Sol2_Parameters}) and 
the eigenvalues of the Schwarzschild metric with $m=1$ and $r_0=3$. 

\subsection{The Buchdahl I  spacetime}
\label{sec:buc}

The Buchdahl spacetime describres a spherically symmetric perfect fluid solution of Einstein equations. The corresponding metric, density, and pressure   read \cite{buchdahl1959general,delgaty1998physical}          
\begin{align}
{\bf g} ^-   =  &   -   A[ (1 + Kr^2)^{3/2} + B(2 - Kr^2)^{1/2}(5 + 2Kr^2) ]^3  \operatorname{d}t \otimes    \operatorname{d}t  \label{eq:S3_Sol_Metric}   \\
&   +  \frac{2(1 + Kr^2)}{2 - Kr^2} \operatorname{d}r \otimes    \operatorname{d}r   
+   r^2   \operatorname{d}\Omega \otimes    \operatorname{d} \Omega \ , \nonumber  \\    
\rho  & =    \frac{3 K (K r^2 + 3)}{16 \pi (K r^2 + 1)^2} \ ,\label{eq:S3_Sol_Energy} \\
p    &  =    - \frac{9K}{16 \pi }  
\frac{B(K r^2 + 1)^{1/2}(2- Kr^2)^{1/2}(2Kr^2 +1) + K^2r^4 -1}{(Kr^2 + 1)^{3/2}[ B(2Kr^2 +5)(2- Kr^2)^{1/2} + (K r^2 +1)^{3/2}]}
\ ,       \label{eq:S3_Sol_Pressure}           
\end{align}        
where   $A$, $B$, and $K$ are  arbitrary constants. On a hypersurface $r=$const., this metric induces the 3-metric 
\begin{align}\label{eq:InducedMetricS3_Sol}
{\mathbf \gamma}^-_{tt} & = -A [  ( Kr^2 +1)^{3/2} + B(2Kr^2 + 5)(2 - Kr^2)^{1/2}]^2 , \\
{\mathbf \gamma}^-_{\theta \theta} &  =  r^2 \ , \quad  {\mathbf \gamma}^-_{\phi \phi}  =  r^2 \sin^2\theta \ , \nonumber
\end{align}      
whose extrinsic curvature is given by
\begin{align}\label{eq:ExtrinicCurvatureS3_Sol} 
{\mathbf K}^- _{tt}                  & =    \frac{3\sqrt{2} K A r (2 - Kr^2)^{1/2}   [ (Kr^2 + 1)^{3/2}  + B(2 - Kr^2)^{1/2}(2Kr^2 + 5)] }{2(1+ Kr^2)^{1/2} (2 - Kr^2)^{1/2}} \\
& \times  [B(1- 2Kr^2) + (1+ Kr^2)^{1/2}(2- Kr^2)^{1/2}]
, \nonumber \\
{\mathbf K}^- _{\theta \theta} & =  \frac{\sqrt{2} (2 - Kr^2)^{1/2}}{2 (Kr^2 + 1)^{1/2}}  \ , \qquad
{\mathbf K}^-_{\phi \phi}        =   \sin^2{\theta}    \, {\mathbf K}^- _{\theta \theta} \ . \nonumber
\end{align}  
Then, imposing the first and second Darmois conditions on a matching hypersurface determined by  a sphere of radius $r= 2 \kappa m$, we obtain that if we choose  the arbitrary constants entering the metric as 
\begin{align}\label{eq:Sol3_Parameters}
A  &  = \frac{(9 \kappa^2 + 12 \kappa  + 4)(\kappa -1)}{108\kappa^2 (3 \kappa -2)} \ ,  \\ 
B  &  = \frac{\kappa^{1/2} (3 \kappa - 4)}{(2 + 3\kappa)( 2\kappa - 2)^{1/2}} \ , \\
K  &  = \frac{1}{2\kappa^2 m^2(3\kappa -2)} \ ,
\end{align}          
then the Darmois matching approach guarantees that the interior Buchdahl I spacetime can be matched with the exterior Schwarzschild spacetime. 

\begin{figure}[h!] 
	\centering
	\subfloat[$p$]{ \includegraphics[width = 0.45\textwidth]{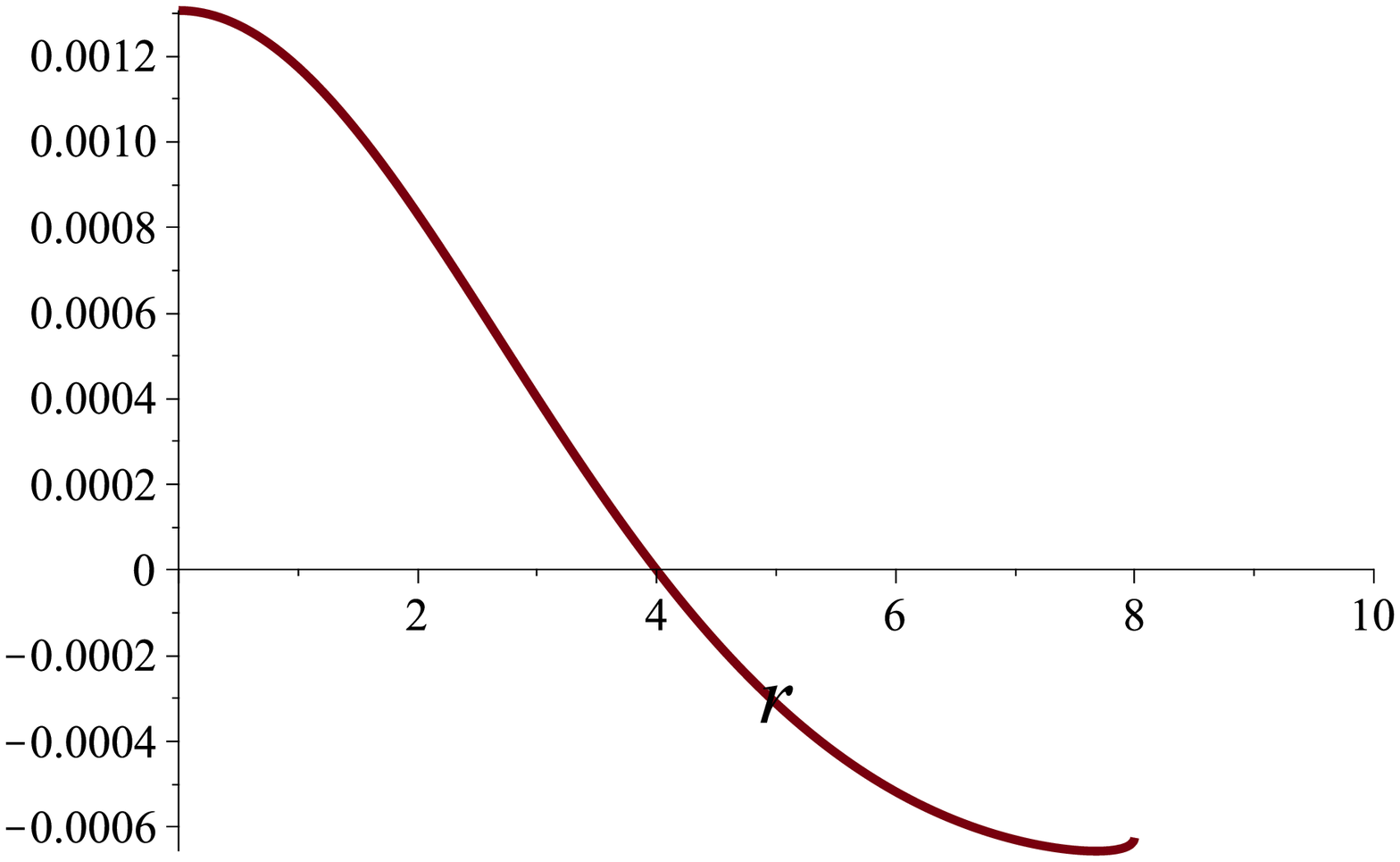}}
	\subfloat[$\rho$]{\includegraphics[width = 0.45\textwidth]{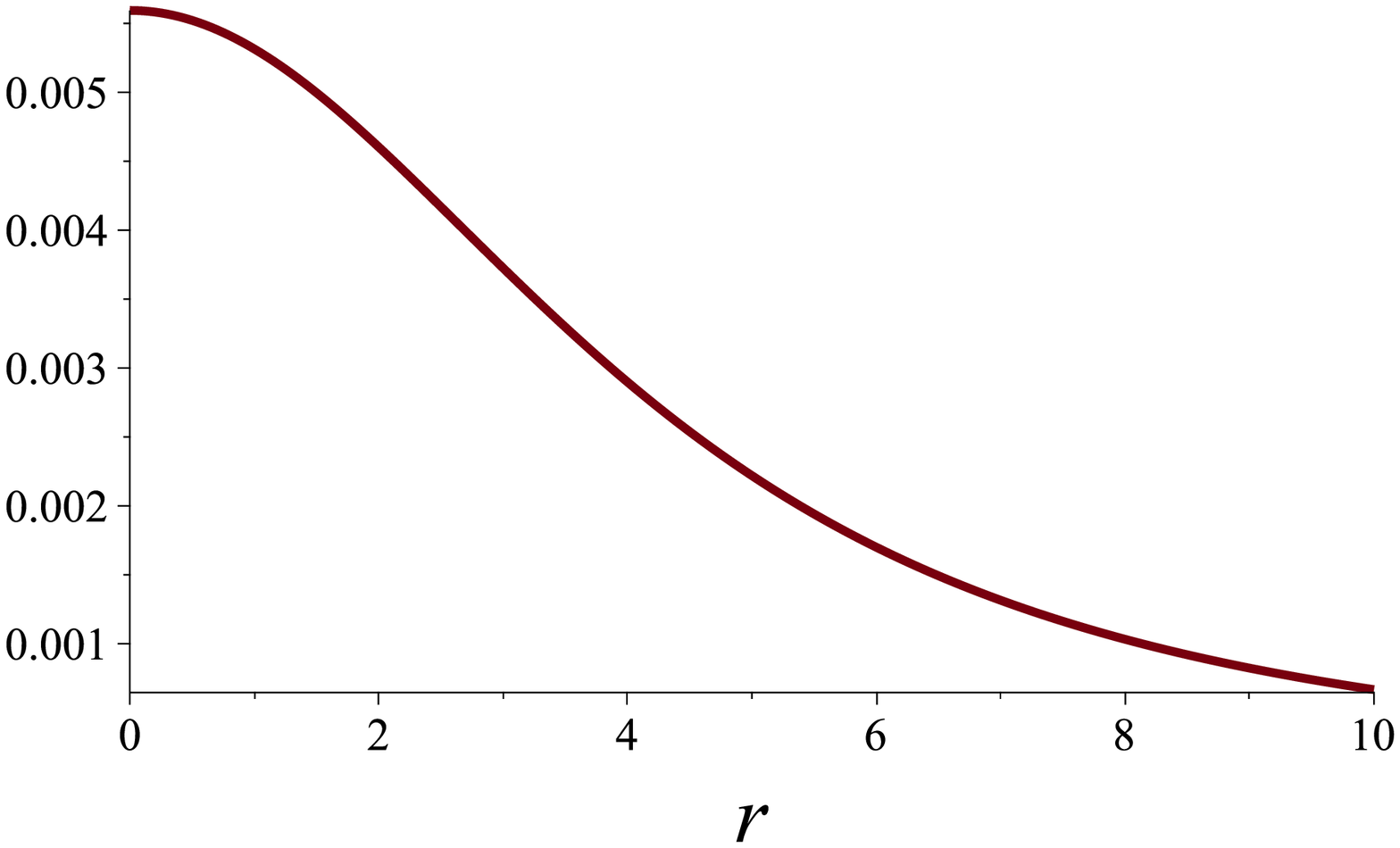}}
	\caption{ The pressure and the energy density of the Buchdahl I spacetime for $\kappa = 2$, $m=1$, and matching radius located at $r=4$. }
	\label{fig:Sol3_Presure_and_Energy} 
\end{figure}

It can also be shown that at the matching radius $r=2\kappa m$, the pressure vanishes identically and the density is nonzero and positive. This behavior is illustrated in Fig. 
\ref{fig:Sol3_Presure_and_Energy}. 

\begin{figure}[h!]
	\centering
	\subfloat[$\lambda_1$]{  \includegraphics[width = 0.5\textwidth]{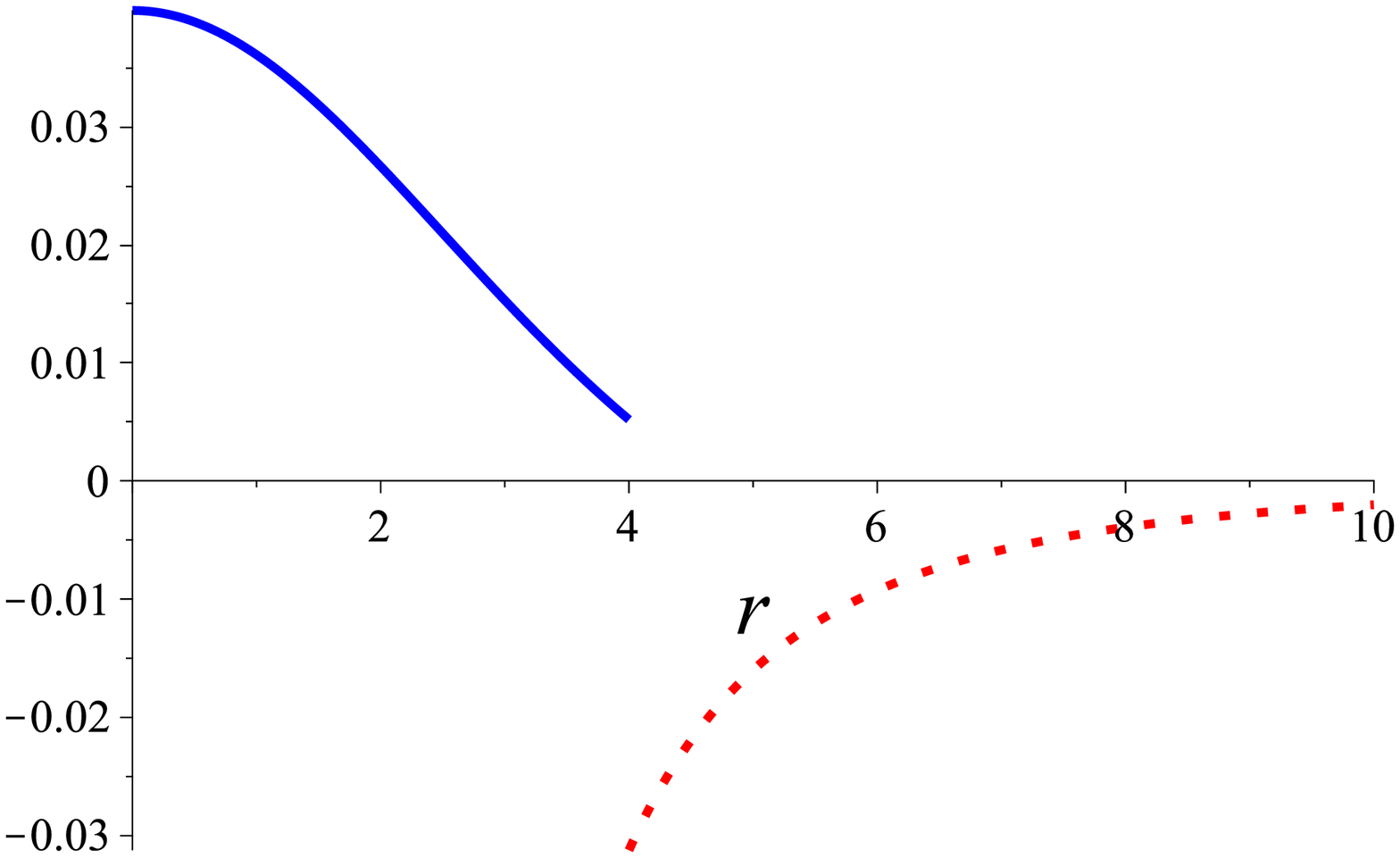} }
	\subfloat[$\lambda_2$]{  \includegraphics[width = 0.5\textwidth]{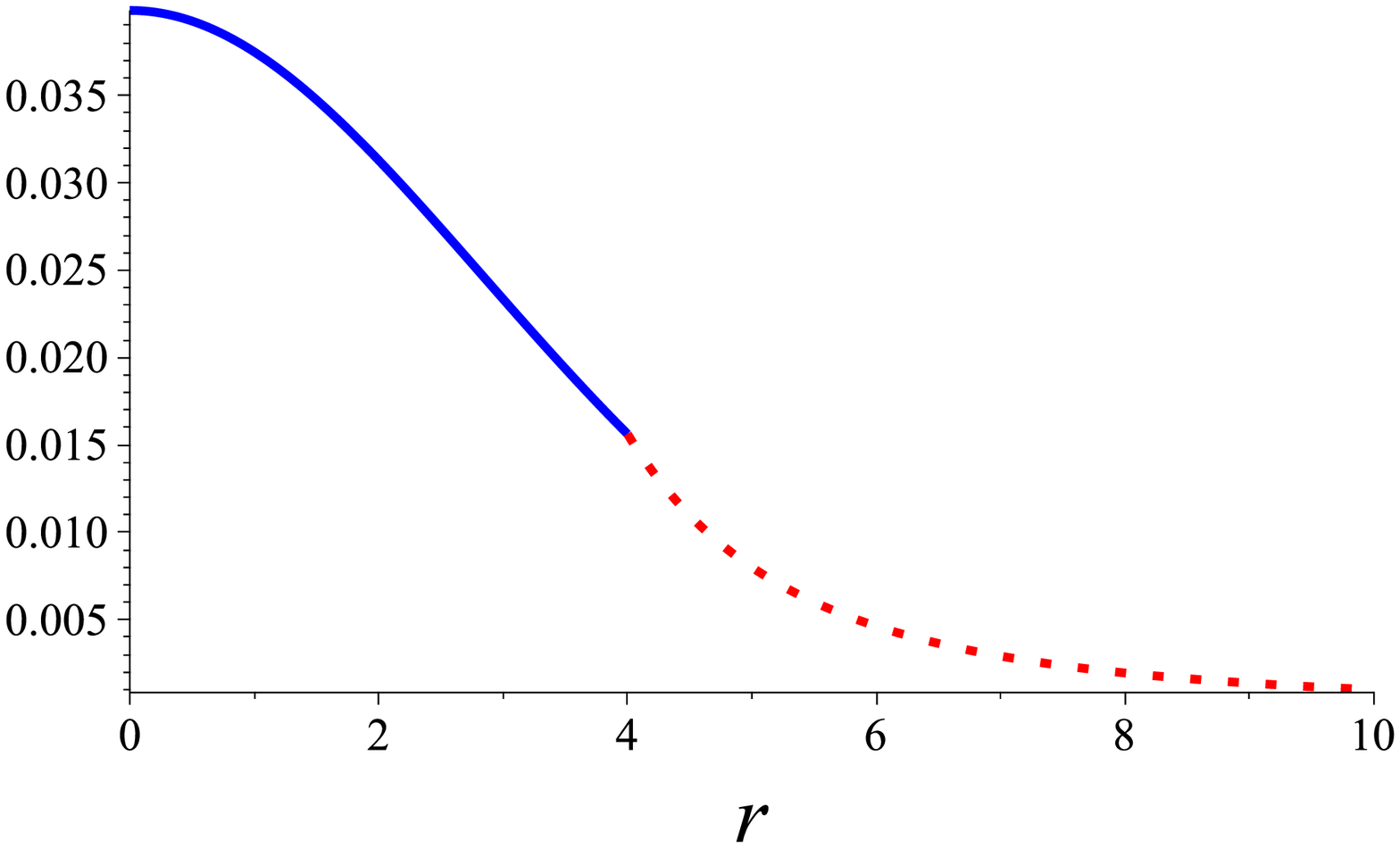} } \\
	\subfloat[$\lambda_3$]{  \includegraphics[width = 0.5\textwidth]{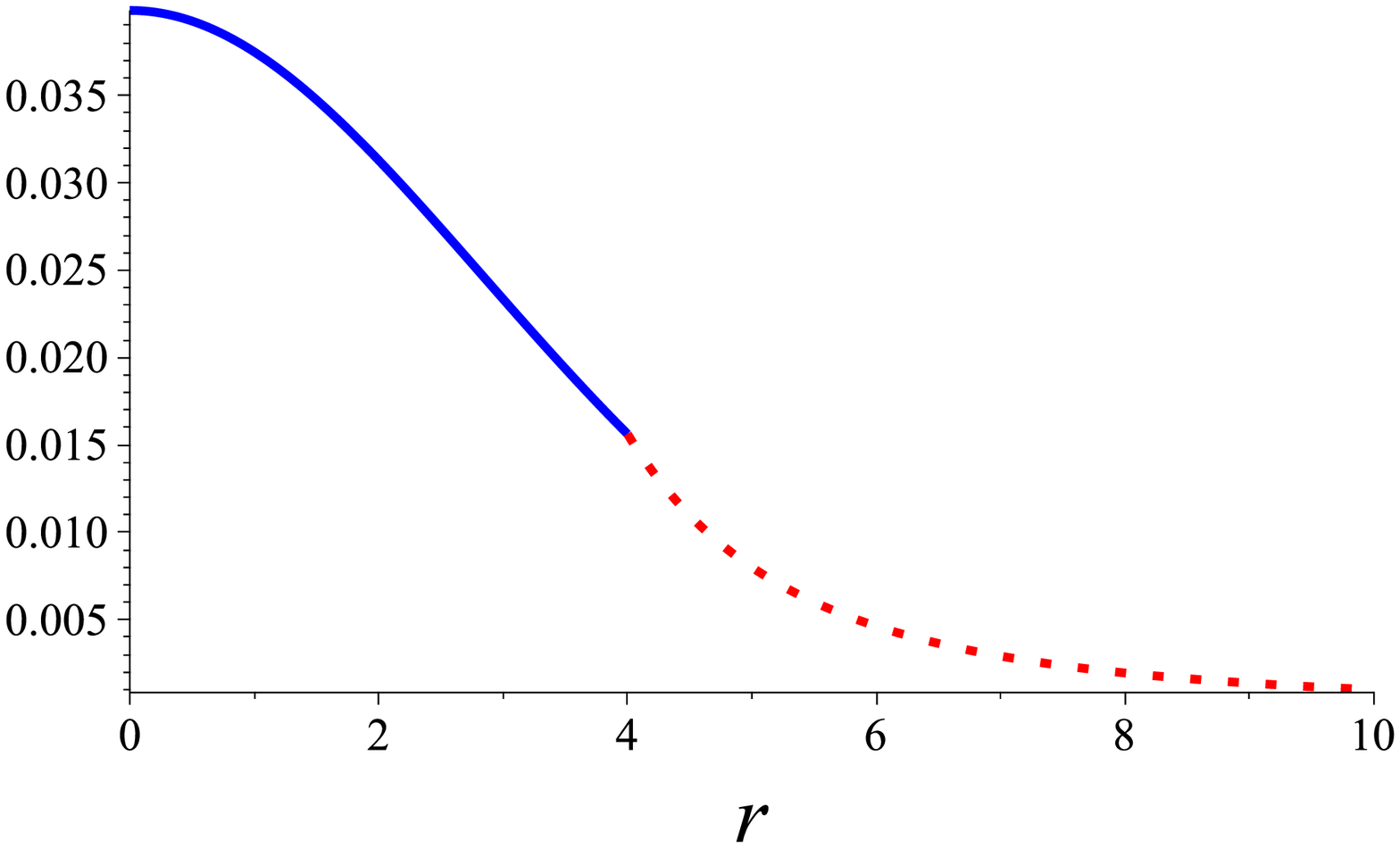} } 
	\subfloat[$\lambda_4$]{  \includegraphics[width = 0.5\textwidth]{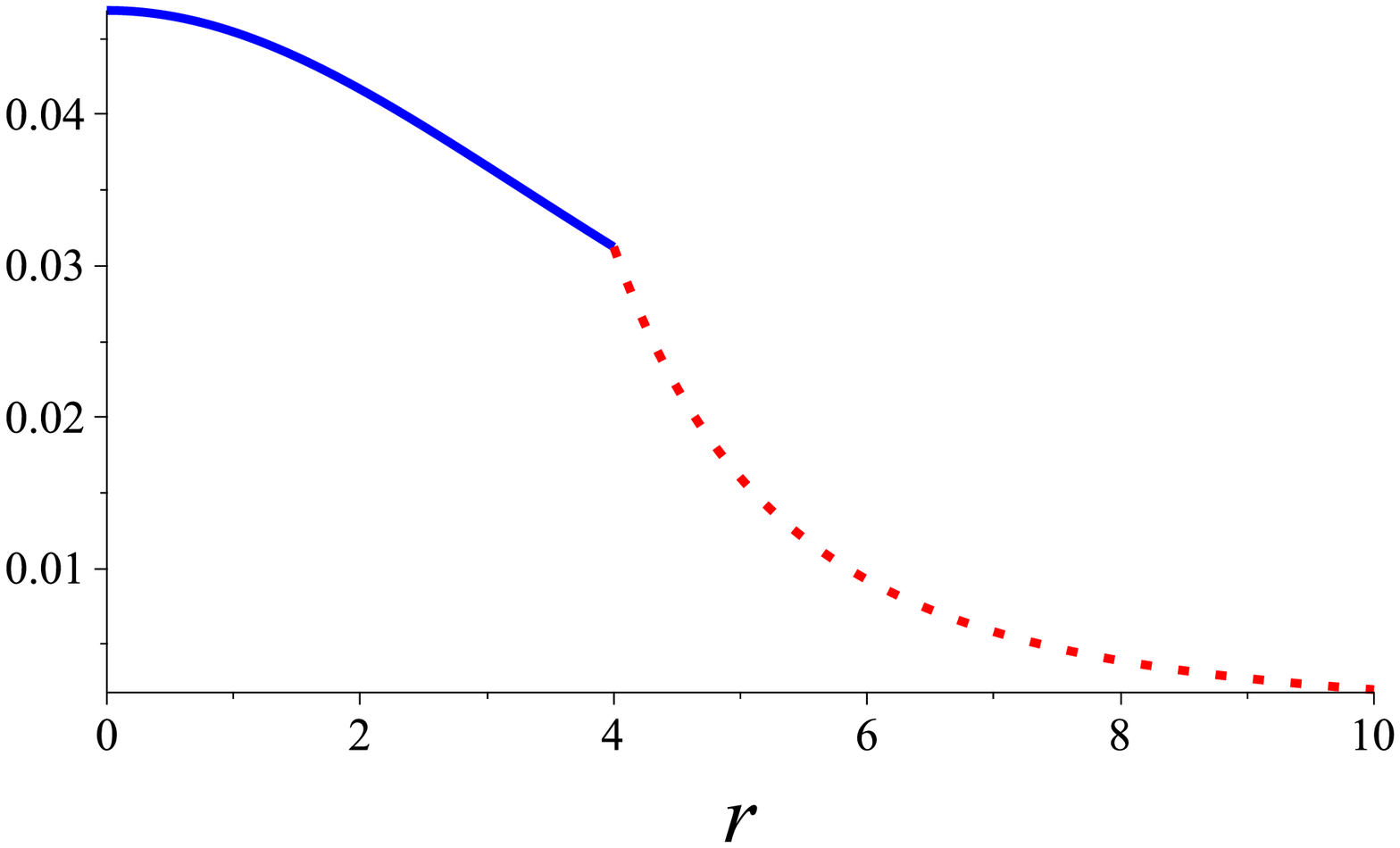} } \\
	\subfloat[$\lambda_5$]{  \includegraphics[width = 0.5\textwidth]{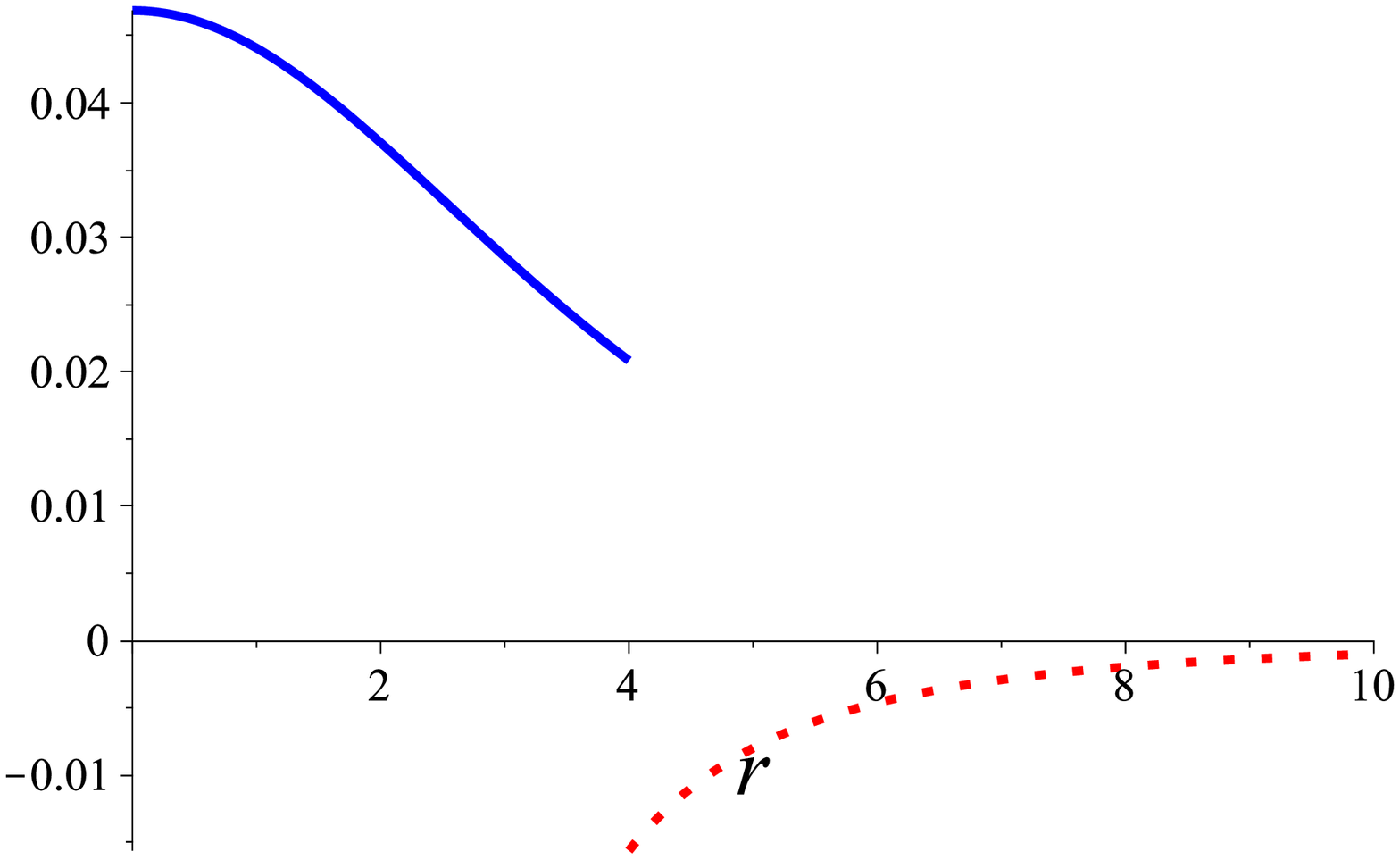} }
	\subfloat[$\lambda_6$]{  \includegraphics[width = 0.5\textwidth]{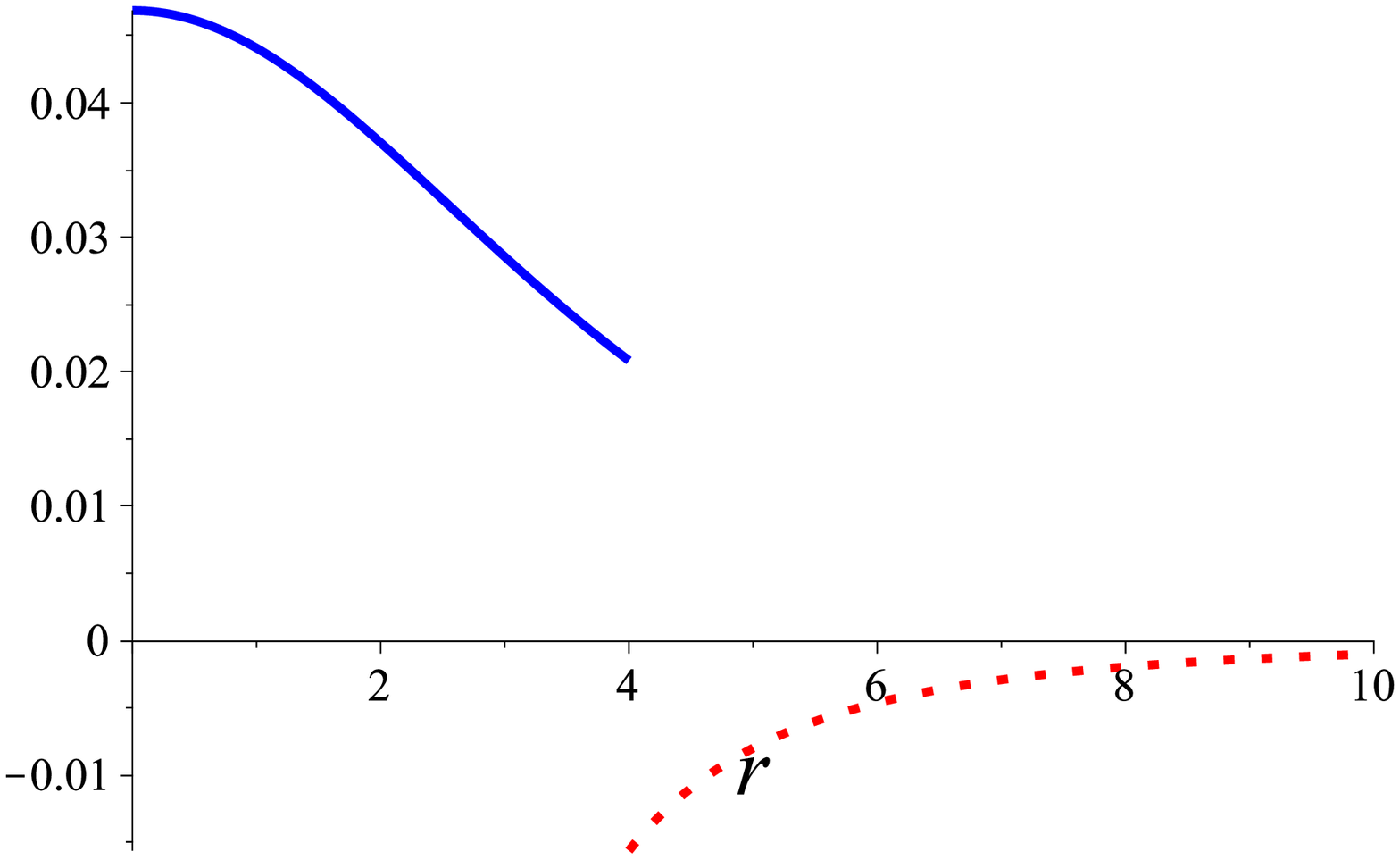} }
	\caption{ The curvature eigenvalues for the interior Buchdahl I metric ($r\leq 4$) and the exterior Schwarzschild metric ($r\geq 4)$ for $\kappa =2$ and $m=1$.}
	\label{fig:Sol3_eigenvalues}
\end{figure}

To carry out  the $C^3$ approach, we choose the orthonormal tetrad as 
\be 
\vartheta^0 = A^{1/2}[ (1 + Kr^2)^{3/2} + B(2 - Kr^2)^{1/2}(5 + 2Kr^2) ]^{3/2}\operatorname{d}t \ ,
\ee
\be
\vartheta^1 =  \left[\frac{2(1 + Kr^2)}{2 - Kr^2}\right]^{1/2} r   \operatorname{d}r\ , \ 
\vartheta^2 =r \operatorname{d} \theta\ ,\ 
\vartheta^3 = r \operatorname{d} \phi \ .
\ee
Then, following the method for calculating the curvature matrix ${\bf R}_{AB}$, we find the following curvature eigenvalues
\begin{align}\label{eq:Sol3_lambda(1)}
\lambda^-_1 =  \frac{3K [2(1-Kr^2)(1+ Kr^2)^{3/2} + B(2- Kr^2)^{1/2}(1- 4K^2r^4 - 6Kr^2)] }
{2(1 + Kr^2)^2 [ (1+ Kr^2)^{3/2}    + B (2 - Kr^2)^{1/2}(5 + 2Kr^2)] } \ ,         
\end{align}    

\begin{align}\label{eq:Sol3_lambda(2-3)}
\lambda^-_2  =  \lambda^-_3 = \frac{3K(2 - Kr^2)^{1/2} [ B(1 - 2Kr^2) + (1 + Kr^2)^{1/2}(2 -Kr^2)^{1/2} ]}
{2(1+Kr^2) [(1+ Kr^2)^{3/2} + B(2 - Kr^2)^{1/2} (5 + 2Kr^2)]}
\end{align}

\begin{align}
\lambda^-_4   & =  -     \lambda^-_1  + 4\pi ( \rho + p ) = \frac{3K}{2(Kr^2 + 1)} \ ,\\
\lambda^-_5  & =  -     \lambda^-_2  + 4\pi ( \rho + p ) = \frac{3K}{2(Kr^2 + 1)^2} \ ,\\
\lambda^-_6   & =  -     \lambda^-_3  + 4\pi (\rho + p ) = \frac{3K}{2(Kr^2 + 1)^2}  \ .
\end{align} 
Then, it can be shown that no solution exists for the $C^3$ matching condition. In fact, the equivalence between the interior and exterior curvature eigenvalues on the matching surface can be reached only for some of the eigenvalues. To illustrate the lack of coincidence, we plot in Fig.  \ref{fig:Sol3_eigenvalues}, the behavior of the eigenvalues as functions of the radial coordinate $r$. For concreteness, the free parameters $A, B$ and  $K$ are chosen as given in Eq.(\ref{eq:Sol3_Parameters}) with  $\kappa = 2$ and $ m=1$.

The above calculations are based upon the use of the explicit definitions of extrinsic curvature and Riemann curvature its eigenvalues. One might wonder whether the curvature eigenvalues are somehow related to components of the extrinsic curvature. The above examples show that there is no such a relationship. In fact, the Riemann tensor determines the curvature of the 4-dimensional spacetime manifold  itself,
${\cal M}$, whereas the extrinsic curvature tensor refers to the curvature of a 3-dimensional submanifold embedded in ${\cal M}$. This implies that the extrinsic curvature depends explicitly on the  embedding. On the other hand, the Riemann tensor is independent of the existence of a submanifold and any embedding. For this reason, the components of the extrinsic curvature are not related to  the eigenvalues of the Riemann tensor. In fact, we will see that their application in the context of the matching procedure leads to contradictory results.

Nevertheless, the extrinsic curvature is extensively used in general relativity to study the evolution of a given spacetime. Indeed, in this case, the embedding is determined by the choice of time and the evolution of the spacetime can be investigated by using the extrinsic curvature of the corresponding 3-dimensional spatial submanifold. This is the subject of numerical relativity \cite{alcubierre2008introduction}.

\subsection{Analysis of the results}
\label{sec:ana}

The results of the previous subsections show that different matching approaches can lead to different results. Whereas according to the Darmois approach, it is possible to match the exterior Schwazschild solution with the Tolman III, Heintzmann II, and Buchdahl I perfect fluid solutions, the $C^3$ matching approach shows that it is not possible.  This seems to indicate that the $C^3$ procedure is more restrictive than the Darmois approach. 

To explain this contradictory result, let us consider the behavior of the thermodynamic variables $p$ and $\rho$ on the matching surface $\Sigma$. In all the three cases investigated above, we can see that the pressure vanishes on $\Sigma$, but the density is different from zero. In the case of the Tolman III perfect fluid, the density is constant everywhere, even outside the body, and in the case of the Heintzmann II and Buchdahl I solutions, the density vanishes only asymptotically at infinity. It seems that for the Darmois matching approach, this physical obstruction is not a problem. In fact, from a physical point of view, one would expect that pressure and density should vanish outside the object. 

On the other hand, the $C^3$ approach can detect the physical obstruction due to a non-vanishing density on the matching surface. In fact, in \cite{gq19}, it was shown that in the case of spherically symmetric perfect fluids, the vanishing of the energy-momentum tensor on the matching surface is a necessary condition to perform the matching procedure. This means that the pressure, as well as the density, should vanish on the matching hypersurface. In this sense, from a physical point of view, we can ensure that the $C^3$ approach is more restrictive than the Darmois approach.

This result also shows that a generalization of Darmois approach would be appropriate to handle the cases in which the jump of the curvature eigenvalues across $\Sigma$ does not vanish. We will propose such a procedure in the next section.


\section{  $C^3$  discontinuous matching}
\label{sec:dis}

The previous results show that the $C^3$ procedure does not allow to match spherically symmetric perfect fluid spacetimes, whose density and pressure are different from zero on the matching surface.  

Now, we will construct a formalism that allows the matching in the case of discontinuities across the matching surface, i.e., 
$\lambda_n^+\neq \lambda_n ^-$ on $\Sigma$ for at least one value of $n$. 
We will use as a conceptual guide Israel's formalism \cite{israel1966singular} that allows the existence of discontinuities of the first and second fundamental forms by introducing an effective energy-momentum tensor on the matching surface $\Sigma$ so that it can be intepreted as a infinitesimal matter shell that join the interior and exterior spacetimes. 
To this end, let us consider the jump of the eigenvalues across $\Sigma$ as 
\be
[\lambda_n] = \lambda_n ^- - \lambda_n ^+ \ ,
\ee
In the case of a matching between an interior perfect fluid solution and the exterior Schwarzschild  vacuum solution, we have shown that the $C^3$ procedure implies that $\rho$ and $p$ should be zero on $\Sigma$. When these conditions are not satisfied, let us define the surface density $\sigma$ and pressure $\pi$ as
\be
\sigma = \rho|_\Sigma\ , \qquad P= p|_\Sigma\ .
\label{bound}
\ee
Then, since in the case of discontinuities we have that $[\lambda_n] \neq 0$, it follows that $\sigma \neq 0$ and $P \neq 0$, in general, as will be shown in concrete examples below. This is equivalent to saying that the explicit values of $[\lambda_n$] should contain information about the physical quantities $\sigma$ and $P$. 
For this reason, we assume that $[\lambda_n]$ is arbitrary in value but finite.
The question is now whether $\sigma$ and $P$ can be used to construct a realistic matter shell on $\Sigma$. To this end, consider the jump of the Einstein tensor on $\Sigma$, i. e., 
\be
[G_{ij}] = G_{ij}^- - G_{ij}^+\ , \quad
G_{ij} ^\pm = \frac{\partial x^\mu_\pm}{\partial \xi^i}
\frac{\partial x^\nu_\pm}{\partial \xi^j} G_{\mu\nu}^\pm\ ,
\ee
where $\xi^i$ are the coordinates of the surface $\Sigma$ and $x^\mu_\pm$ are the coordinates of the interior and exterior spacetimes, respectively. Then, $G_{ij}^\pm$ is the Einstein tensor induced on $\Sigma$. Furthermore, we introduce an energy-momentum tensor $S_{ij}$ on $\Sigma$ as
\be 
[G_{ij}] = k S_{ij}\ .
\label{ie1}
\ee
Certainly, it is always possible to introduce algebraically an energy-momentum tensor in this way. However, the essential point is whether $S_{ij}$ is physically meaningful. To guarantee the fulfillment of this condition, we demand that $S_{ij}$ be induced by the energy-momentum tensors of the interior and exterior spacetimes and be in agreement with their physical significance. Then,  in the case of the  perfect fluid we are considering here, we demand that 
\be
S_{ij} = [T_{ij}] = T_{ij}^--T_{ij}^+ =  (\sigma+P)u_iu_j + P \gamma_{ij}\ ,
\label{ie2}
\ee
where $T^\pm_{ij}$ are the energy-momentum tensors and $\gamma_{ij}=\gamma_{ij}^\pm$ is the metric tensor induced on $\Sigma$, respectively.

In summary, in the case of discontinuities, we will say that an interior spacetime can be matched with an exterior one along a boundary shell located on $\Sigma$, if there exist a density $\sigma$ and a pressure $P$, satisfying the induced Einstein equations (\ref{ie1}) and (\ref{ie2}) and the boundary condition (\ref{bound}). 

To test the above procedure, we consider now the explicit examples presented in the previous sections. First, we notice that in the case of spherical symmetry the coordinates on both sides of the boundary can be chosen as $x^\mu_\pm = (t,r,\theta,\phi)$ and on the matching surface as $\xi^i =(t,\theta,\phi)$. Then, all the components of the quantities $\frac{\partial x^\mu}{\partial \xi^i}$ are constant and the induced tensors can be calculated in a straightforward way.

{\bf Tolman III}

In this case, from the results presented in Sec. \ref{sec:tol}, we obtain for the jump of the eigenvalues along the matching surface $r=K$ the following expressions
\be 
[\lambda_2]= [\lambda_3]= [\lambda_4]=0\ , \quad
[\lambda_1]= [\lambda_5]= [\lambda_6]= \frac{3m}{K^3} = 4\pi \sigma \ ,
\ee
which agrees with the result that on the matching surface the pressure vanishes. Furthermore, the jump of components of the induced Einstein tensor can be expressed as
\be
[G_{tt}] = \frac{4mr^2 + K^3 - K^3f^2(r) }{K^3 r^2} \gamma_{tt} \ ,
\ee
\be
[G_{\theta\theta}] =  \frac{6mr^2}{K^3} \, \frac{f(K)-f(r)}{3f(K) - f(r)}\ , \quad
[G_{\phi\phi}]= \sin^2\theta \, [G_{\theta\theta}]\ .
\ee
It is then easy to see that on the matching surface $r = K$, the induced Einstein equations for dust are satisfied 
\be
[G_{ij}] = k \sigma u_i u_j\ , \quad u^i=(-1,0,0)\ ,
\ee
proving that, in fact, a realistic dust shell can be introduced that allows us to match, in the framework of the $C^3$ matching procedure, the interior Tolman III solution with the exterior Schwarzschild spacetime.

An explicit calculation gives
\be
[G_{tt}] = \frac{6m}{K^3} \ ,
\ee
\be
[G_{\theta\theta}] =  
[G_{\phi\phi}] = 0 .
\ee

{\bf Heintzmann II} 

In this case, from the results presented in Sec. \ref{sec:hei}, we obtain for the jump of the eigenvalues along the matching surface $r=r_0$ the following expressions
\be 
[\lambda_2]= [\lambda_3]= [\lambda_4]=0\ , \quad
[\lambda_1]= [\lambda_5]= [\lambda_6]=  \frac{(14 m - 9r_0)m}{3r_0^3(m - r_0)}  =4\pi \sigma \ .
\ee
From the expressions for the induced Einstein tensor, we obtain the jump
\be
[G_{tt}] = \frac{2 (14 m - 9r_0)m}{3r_0^3(m - r_0)} \ ,
\ee
\be
[G_{\theta\theta}] =  
[G_{\phi\phi}] = 0\ .
\ee
In this case, the induced Einstein equations $[G_{ij}] = k \sigma u_i u_j$ are satisfied for
\be
\sigma = \frac{(14 m - 9r_0)m}{12 \pi r_0^3(m - r_0)} \ ,  \quad u^i=(-1,0,0) \ , 
\ee
an expression that fulfills the compatibility condition (\ref{bound}).

{\bf Buchdahl I}

From the results presented in Sec. \ref{sec:buc}, we obtain the following jumps for the eigenvalues along the surface $r = 2 \kappa m$ 
\be 
[\lambda_2]= [\lambda_3]= [\lambda_4]=0\ , \quad
[\lambda_1]= [\lambda_5]= [\lambda_6]=  \frac{9 \kappa - 4}{24\kappa^4 m^2}  = 4\pi \sigma \ .
\ee
Furthermore, the jump of the induced Einstein tensor reads
\be
[G_{tt}] = \frac{9 \kappa - 4}{12\kappa^4 m^2} \ ,
\ee
\be
[G_{\theta\theta}] = 
[G_{\phi\phi}] = 0 \ .
\ee
Then, the induced Einstein equations $[G_{ij}] = k  \sigma u_i u_j$ are satisfied for
\be
\sigma =  \frac{9 \kappa - 4}{96 \pi \kappa^4 m^2} \ ,  \quad u^i=(-1,0,0) \ ,
\ee
in accordance with the compatibility condition (\ref{bound}).

Notice that in all the above examples the energy-momentum tensor of the dust shell can be expressed as 
\be
S_{ij} = 2 [\lambda_1] u_iu_j \ ,
\ee
indicating that the properties of the boundary shell are determined in an invariant manner by the curvature eigenvalues.

\section{Discussion and remarks}
\label{sec:con}

In this work, we have analyzed the problem of matching exact solutions of Einstein equations in order to describe a spacetime completely. We limit ourselves to the case of spherically symmetric solutions of Einstein equations. Since the vacuum Schwarzschild solution is singular at the origin of coordinates, it is believed that an appropriate non-vacuum and singularity-free solution can be used to ``cover" the Schwarzschild singularity in such a way that the entire spacetime is regular. This is how one expects that classical general relativity can get rid of curvature singularities. This seems to be a simple method to solve such an important problem of general relativity. However, the problem arises of matching the non-vacuum and vacuum solutions in such a way that the entire differential manifold is well behaved. This is why the matching problem is very important in general relativity.

We have applied the Darmois and the $C^3$ approaches to three different perfect solutions of Einstein equations, which could be considered as appropriate interior candidates to be matched with the exterior Schwarzschild solution. These are the Tolman III, Heintzmann II, and Buchdahl I solutions. We have shown that  the mentioned matching procedures lead to contradictory results. According to Darmois approach, all the three candidates satisfy the matching conditions and can be interpreted as interior counterparts of the exterior Schwarzschild metric. However, the $C^3$ approach shows that none of the three solutions satisfy the $C^3$ matching conditions. We explain this contradictory result by noticing that in all three perfect fluid solutions, the energy density shows a discontinuity across the matching surface. Although this is not an obstacle for the Darmois procedure, the $C^3$ approach demands that the pressure and the density as well vanish on the matching surface. This is why both approaches lead to different results. We mention that we obtained the same result in the case of the Durgapal IV and V spacetimes \cite{durgapal1982class, delgaty1998physical}, which are spherically symmetric perfect fluid solutions of Einstein equations.

To handle the case in which discontinuities are present along the matching surface $\Sigma$, we propose in this work a generalization of the $C^3$ matching procedure. It consists essentially on demanding that the 3-dimensional hypersurface $\Sigma$ be also described by a solution of Einstein equations. In fact, we consider the induced Einstein tensor on $\Sigma$ and show that it can be represented as a realistic energy-momentum tensor that describes the matter inside a boundary shell located on $\Sigma$.
In the cases considered in this work, it turned out that the boundary corresponds to a dust shell. For more general interior solutions, we expect to obtain shells with more intricate internal structures.

\section*{Acknowledgments}  A.C.G-P. is thankful to the Departamento de Gravitaci\'on y Teor\'ia de Campos (ICN-UNAM) for its hospitality during his research fellowship.  
Also,  A.C.G-P would like to GTD-ICN team for useful comments and discussions. This work was partially supported by the Programa Capital Semilla para Investigaci\'on, 
Proyecto 2490, Programa de Movilidad Académica, VIE-UIS, 
UNAM-DGAPA-PAPIIT, Grant No. 114520, and
Conacyt-Mexico, Grant No. A1-S-31269.

\end{document}